\documentclass[a4paper]{amsart}

\usepackage{amssymb}
\usepackage{amsmath}
\usepackage{graphicx}
\usepackage{enumerate}
\usepackage{eurosym}
\usepackage{hyperref}
\usepackage[ansinew]{inputenc}
\usepackage{verbatim}
\usepackage{color}

\DeclareMathOperator*\esssup{ess\,sup}
\DeclareMathOperator*\essinf{ess\,inf}

\newcommand{\A}{\mathcal{A}}

\newcommand{\E}{\mathcal{E}}
\newcommand{\F}{\mathcal{F}}
\newcommand{\G}{\mathcal{G}}
\newcommand{\Hcal}{\mathcal{H}}

\newcommand{\M}{\mathcal{M}}

\newcommand{\Q}{\mathcal{Q}}

\newcommand{\X}{\mathcal{X}}

\newcommand{\EE}{\mathbb{E}}
\newcommand{\FF}{\mathbb{F}}
\newcommand{\GG}{\mathbb{G}}
\newcommand{\HH}{\mathbb{H}}

\newcommand{\NN}{\mathbb{N}}

\newcommand{\RR}{\mathbb{R}}

\newcommand{\comments}[1]{}

\title[Utility Maximization and Indifference Value under Risk and Incomplete Constraints]{Utility Maximization and Indifference Value under Risk and Information Constraints for a Market with a Change Point}
\author{Oliver Janke}
\address{Humboldt-Universit\"{a}t zu Berlin, Department of Mathematics, Unter den Linden 6, 10099 Berlin, Germany.} 
\email{janke@math.hu-berlin.de} 
\date{October 27th, 2016}
\keywords{Utility maximization, risk constraint, incomplete information, random time, change point, martingale method, enlargement of filtrations, value of information}
\subjclass[2010]{{Primary 91G10, 94A17}; Secondary{ 60G40, 60G44, 91B25, 91B70}}
\thanks{The author thanks Ulrich Horst, Qinghua Li and Anna-Maria Hamm for helpful comments and suggestions.}

\begin{document}

\newtheorem{theorem}{Theorem}[section]
\newtheorem{proposition}[theorem]{Proposition}
\newtheorem{lemma}[theorem]{Lemma}
\newtheorem{definition}[theorem]{Definition}
\newtheorem{definition theorem}[theorem]{Definition and Theorem}
\newtheorem{remark}[theorem]{Remark}
\newtheorem{example}[theorem]{Example}
\newtheorem{examples}[theorem]{Examples}
\newtheorem{corollary}[theorem]{Corollary}
\newtheorem{assumption}[theorem]{Assumption}
\newtheorem{problem}[theorem]{Problem}

\begin{abstract}
In this article we consider an optimization problem of expected utility maximization of continuous-time trading in a financial market. 
This trading is constrained by a benchmark for a utility-based shortfall risk measure. 
The market consists of one asset whose price process is modeled by a Geometric Brownian motion where the market parameters change at a random time. 
The information flow is modeled by initially and progressively enlarged filtrations which represent the knowledge about the price process, the Brownian motion and the random time. 
We solve the maximization problem and give the optimal terminal wealth depending on these different filtrations for general utility functions by using martingale representation results for the corresponding filtration. 
Moreover, for a special utility function and risk measure we calculate the utility indifference value which measures the gain of further information for the investor. 
\end{abstract}

\maketitle

\section{Introduction}

For dealing with utility maximization problem there are in general two methods: the {\em dynamic programming} approach 
and the duality or {\em martingale approach} which was used by Kramkov \& Schachermayer \cite{Kramkov1999}, 
besides others, in a general incomplete semimartingale model. 
The main idea of the martingale method which we use here is to obtain the optimal terminal wealth and find the corresponding optimal value process by representing it as a martingale under some equivalent measure. 
Such representation theorems were discussed by, among others, Bielecki \& Rutkowski \cite{Bielecki2002}, Callegaro et al.\ \cite{Callegaro2013}, and Jeanblanc \& Song \cite{Jeanblanc2015}, see also the references therein. \\

A resent work by Fontana et al.\ \cite{Fontana2013} established martingale representation results in a general class of continuous asset price models in which the drift, the volatility as well as the driving Brownian motion change at a random time which is not necessarily a stopping time. 
Inspired by this model, we consider in this paper a utility maximization problem under a shortfall risk constraint under incomplete information in a continuous-time Black-Scholes market but which consists of only one risky asset driven by one Brownian motion. 
So in contrast to other authors we deal with the case where the drift as well as the volatility process change at a random time and use several filtrations to model the information gap of the investor: 
they represent the knowledge of the price process or the underlying Brownian motion up to the present time as well as the knowledge about the occurrence time of the random change point. 
Therefore, this paper also deals with initially enlarged filtrations. 
\cite{Fontana2013} established martingale representation results for these filtrations which we will use to solve the optimization problem. \\
We assume the typical Lipschitz conditions for the price process and consider general utility functions defined on the positive semi-axis which satisfy the Inada conditions as well as utility-based risk measures modeled by convex loss functions. 
Assuming the existence of a Brownian motion and of a minimal equivalent local martingale measure w.r.t.\ the given filtration we first represent the optimal contingent claim which has to be hedged. 
Depending on the underlying filtration the markets are complete in the sense that any contingent claim can be hedged by a self-financing trading strategy, or incomplete where the corresponding trading strategy is no longer self-financing, but mean-self-financing and risk-minimizing in the sense of \cite{Foellmer1986}. \\
Moreover, we compare the gain of the additional information for the investor: we calculate the utility indifference value between a smaller and a larger filtration, i.\,e.\ the largest price an investor is willing to pay for additional information such that his final output is the same under both information flows. 
For initially enlarged filtration, this value was considered by \cite{Amendinger2003} for power, logarithmic and exponential utility functions. \\

Portfolio optimization is a central field of modern financial mathematics. 
Beginning with the work of Merton \cite{merton} who solved the utility maximization problem from terminal wealth for power, logarithmic and exponential utility functions many authors considered similar problems in different mathematical models of the financial markets. 
A new branch of research was founded as Artzner et al.\ \cite{Artzner1999} mathematically defined measures of risk, which were then further developed by, for example, F\"{o}llmer \& Schied \cite{Foellmer2002}.
With this new indicator of a financial product, the question of portfolio optimization under risk constraints has been an active topic of research.
Financial crises in the past decade raised even more alert to risks resulted from portfolio strategies.\\
A special risk measure, namely the shortfall risk, was widely studied by many authors, e.\,g.\  Leibowitz \& Henriksson \cite{Leibowitz1989}, Rockafellar \& Uryasev \cite{Rockafellar2000}, Acerbi \& Tasche \cite{Acerbi2002}, Bertsimas et al.\ \cite{Bertsimas2004} and Goldberg et al.\ \cite{Goldberg2013}, to name some of them. \\ 
Janke \& Li \cite{Janke2016}, among other authors, recently solved the utility maximization problem under a shortfall risk constraint in a general semimartingale framework. 
They showed that under mild assumptions on the utility function and the loss function, that the Lagrange function is a usual utility function whose asymptotic elasticity is less than one. An unconstrained maximization problem where the utility is the Lagrange function can then be solved via the duality approach introduced by Kramkov \& Schachermayer \cite{Kramkov1999}. \\
In recent years, portfolio optimization was considered under the additional topic of incomplete information. 
In contrast to former works, it means that the portfolio manager does not have full access to all information of the financial market, e.\,g.\ he only observes the stock prices but does not exactly know the drift and volatility parameters, which seems to be a more realistic approach. 
Mathematically, this is modeled by different filtrations generated by e.\,g. the price process which represent the information flow of the agent. 
His chosen wealth process must be adapted to the given filtration which is in general smaller than the filtration generated by the underlying Brownian motion. \\ 
Lakner \cite{Lakner1995} solved the utility maximization problem from consumption as well as from terminal wealth under partial information in a Black-Scholes financial market. 
Here, the underlying Brownian motion and the drift process are not known by the investor whose trading decisions must be adapted to the natural filtration of the price process. 
Later in \cite{Lakner1998}, he modeled the drift by a Gaussian process and derived the optimal trading strategy in this case. \\
Similar models were also considered by other authors:
Hahn et al.\ \cite{Hahn2007} considered the Hobson-Rogers model for the volatility. 
Hereby, the investor maximizes his expected utility only in the knowledge of the stock prices. 
By means of Malliavin calculus they derived an explicit formula for the optimal trading strategy and applied their results to market data. \\
Frey et al.\ \cite{Frey2012} as well as Nagai \& Runggaldier \cite{Nagai2008} solved a utility maximization problem for logarithmic and power utilities and determined a complete observation problem for this partially observable problem based on unnormalized and normalized filter values. 
Via dynamic programming approach they obtained a nonlinear PDE and derived a unique viscosity solution which can be in general computed by Monte Carlo simulation.
Following the same approach, Capponi et al.\ \cite{Capponi2015} used portfolios consisting of one stock, one defaultable bond and a bank account. 
The partial observed problem was reduced to a risk-sensitive control problem with full observation. 
Splitting up this problem into two subproblems, they obtained two coupled HJB equations for the optimal value function. \\ 
Mania \& Santacroce \cite{Mania2010} as well as Covello \& Santacroce \cite{Covello2010} considered an optimization problem for power and exponential utility functions where the investor has only information about the price process. 
The related value process is formulated in terms of observable processes and is then characterized as the unique semimartingale backward stochastic differential solution. 
Amendinger \cite{amendinger} and Amendinger et al.\ \cite{Amendinger2003} considered a utility maximization problem but where the investor has additional information already at time zero which is modeled by an initially enlarged filtration. 
They used martingale representation theorems for this filtration to solve the optimization problem. \\
Related utility maximization problems under a risk constraint with incomplete information were considered by Gabih \cite{Gabih2005} and Rudloff et al.\ \cite{Rudloff2008}.
Both used utility-based shortfall risk as risk measure and, corresponding to \cite{Lakner1995, Lakner1998}, assumed an unknown Brownian motion and drift process, which is modeled by a finite-state Markov chain. 
They solved the problem and derived the optimal trading strategies by using Malliavin calculus. \\

Compared to existing works on this topic we consider the utility maximization problem under a risk constraint and under different information desks which were not covered so far in the literature to the best of our knowledge. 
Therefore, we fill the gap between the previous works about utility maximization 
\begin{enumerate}
\item[-] under incomplete information without (e.\,g.\ \cite{Lakner1995}) or with initially enlarged filtrations (e.\,g.\ \cite{Amendinger2003}), 
\item[-] with a risk constraint (e.\,g.\ \cite{Janke2016}), 
\item[-] with risk constraint and unknown drift parameter (e.\,g.\ \cite{Rudloff2008}) 
\item[-] and utility indifference pricing for initially enlarged filtrations with common utility functions (e.\,g.\ \cite{Amendinger2003})
\end{enumerate}
and martingale representation results for several filtrations where only the price process is known and the market parameter change at a random time (e.\,g.\ \cite{Fontana2013}). \\

The remainder of the paper is as follows:
In Section \ref{sec model} we introduce the financial market, the considered filtrations and give the most important definitions for no arbitrage, the utility function as well as the risk measure. 
Next in the main Section \ref{sec solution}, we first formulate the general constrained optimization problem and derive the unconstrained auxiliary problem (Subsection \ref{subsec general problem}) and then solve this problem for different filtrations, first for initially enlarged filtrations (Subsection \ref{subsec g w tau}), then for progressively enlarged filtrations (Subsection \ref{subsec g w}), and last for the price process filtration (Subsection \ref{subsec f s}). 
After that, we calculate the utility indifference value for a special utility and loss function to measure the gain of additional information for the investor (Subsection \ref{subsec uiv}).  
We conclude the paper in Section \ref{sec conclusion}.

\section{The model} \label{sec model}

Let $(\Omega, \mathcal{A}, P)$ be a given probability space and let $T \in (0,+\infty)$ be a fixed time horizon. Let us assume that all random variables and stochastic processes which we will introduce in this paper, are measurable w.\,r.\,t.\ the $\sigma$-algebras $\mathcal{A}$ and $\mathcal{A} \otimes \mathcal{B}([0,T])$, respectively. 
Let $\mathbb{A}=(\mathcal{A}(t))_{t \in [0,T]}$ be a filtration on $(\Omega, \mathcal{A},P)$ which is assumed to be right-continuous and $P$-complete and which is specified later. 
Let us introduce a random change point in the market, i.\,e.\ a random time between $0$ and $T$ where the market parameters will change. 
It is represented by a random time $\tau:\Omega \to \RR_+$, which is an $\mathcal{A}$-measurable random variable and not necessarily an $\mathbb{A}$-stopping time, cf.\ \cite{Fontana2013} and the model therein. \\
Moreover, let $W=(W_t)_{t \in [0,T]}$ be a one-dimensional Brownian motion on $(\Omega,\mathcal{A},P)$. \\
Let us now consider a financial market with one bond (riskfree) and one stock (risky). 
The setting of only two stocks are just made due to simplicity. 
We assume that the interest rate $r=0$ and that the (discounted) price process of the stock follows a geometric Brownian motion and is given by
$$ S \ = \ S_0 \, \mathcal{E}(\tilde{S}), $$
where $\mathcal{E}(\tilde{S})=\exp(\tilde{S} - \langle \tilde{S},\tilde{S} \rangle /2)$ denotes the stochastic exponential. 
The process $\tilde{S}=(\tilde{S}_t)_{t \in [0,T]}$ is given by:
\begin{equation} \label{eq tilde s general} 
\left\{ \begin{array}{rcl} d \tilde{S}_t & = &  \left( {\bf 1}_{\{t \leq \tau\}} \mu^1(t,\tilde{S}(t)) + {\bf 1}_{\{t > \tau\}}\mu^2(t,\tilde{S}(t)) \right) \, dt + \left( {\bf 1}_{\{t \leq \tau\}} \sigma^1(t,\tilde{S}(t)) + {\bf 1}_{\{t > \tau\}}\sigma^2(t,\tilde{S}(t)) \right) \, dW_t, \\
\tilde{S}_0 & = & 0. \end{array} \right.
\end{equation}
Let us first state some technical assumptions about the Lipschitz conditions of the drift and volatility processes. 

\begin{assumption} \label{ass 1}
For $i=1,2$ the functions $\mu^i:[0,T] \times \RR \to \RR^*$ and $\sigma^i:[0,T] \times \RR \to (0,+\infty)$ are Borel-measurable functions 
and satisfy:
\begin{enumerate}
\item[(a)] There exists a constant $K >0$ such that:
\begin{equation*}
|\mu^i(t,x) - \mu^i(t,y)| \ \leq \ K|x-y|, \quad |\sigma^i(t,x) - \sigma^i(t,y)| \ \leq \ K|x-y|,
\end{equation*}
for all $t \in [0,T]$, for all $x,y \in \RR$, $i=1,2$. 
\item[(b)] The function $(t,x) \mapsto \sigma^i(t,x)$ is jointly continuous in $(t,x) \in [0,T] \times \RR$, for $i=1,2$. 
\end{enumerate}
\end{assumption}

$\mathbb{A}=(\mathcal{A}_t)_{t \in [0,T]}$ describes the generic filtration on $(\Omega,\mathcal{A},P)$ w.\,r.\,t.\ which the process $S$ is a semimartingale. 
Moreover, let us define troughout the paper:
\begin{eqnarray} \label{eq sigma general} 
\sigma_{t} & := & {\bf 1}_{\{t \leq \tau\}} \sigma^{1}(t,\tilde{S}_{t}) + {\bf 1}_{\{t > \tau\}}  \sigma^{2}(t,\tilde{S}_{t}), \qquad 0 \leq t \leq T. 
\end{eqnarray}

\begin{definition} \label{def equivalent martingale measure}
A probability measure $Q$ which satisfies
\begin{itemize}
\item[(i)] $Q \sim P$, i.e. $P(A)=0$ if and only if $Q(A)=0$ ($P \ll Q \ll P$) for any measurable set $A$,
\item[(ii)] the price process $S$ is a local martingale under $Q$
\end{itemize}
is called {\em equivalent local martingale measure}. 
The set of all equivalent local martingale measures is denoted by $\mathcal{Q}^e$.
\end{definition}

For dealing with incomplete information of the financial market, let us consider different filtrations in this paper.
They represent the knowledge of the Brownian motion $W$ or the price process $S$ up to the present time as well as the knowledge of the random time $\tau$. 
They are rigorously defined in the several subsections. \\
Let us denote by $\FF^W=(\F^W_t)_{t \in [0,T]}$ and $\FF^S=(\F^S_t)_{t \in [0,T]}$ the filtrations which are generated by $W$ and $S$, respectively, i.\,e.\ $\mathcal{F}^W_t:=\sigma(W(s): s \leq t)$ represents the knowledge of the Brownian motion and $\mathcal{F}^S_t:=\sigma(S(s): s \leq t)$ represents the knowledge of the price process up to time $t$.
Moreover, for general issues we use the filtration $\FF=(\F_t)_{t \in [0,T]}$ which is some element out of the set of the considered filtrations. 
Furthermore, let $L(S,\mathbb{F})$ denote the set of all $S$-integrable $\mathbb{F}$-predictable processes. 
The family of all $\mathbb{F}$-local martingales is denoted by $\mathcal{M}_{loc}(\mathbb{F})$. \\

Trading in a financial market is often connected with the assumption of no arbitrage, which intuitively describes that an investor cannot make profit for sure without a risk. There are different definitions of no arbitrage properties. In this paper, we will focus on two of them. 

\begin{definition}[No arbitrage] {\em (cf.\ 
\cite{Kardaras2012, Delbaen1994})}
\begin{enumerate}
\item[(a)] A nonnegative $\mathcal{F}_T$-measurable random variable $\zeta$ is said to yield an Arbitrage of the First Kind in $\mathbb{F}$ if $P(\zeta > 0) > 0$ and if for any $x \in (0,+\infty)$ there exists an element $h^x \in L(S,\mathbb{F})$ such that $x + \int h^x dS \geq 0$, $P$-a.s., and $x + \int_0^T h^x_t dS_t \geq \zeta$, $P$-a.s. If there does not exist such a random variable we say that the {\em No Arbitrage of the First Kind {\bf (NA1)}} condition holds for $\mathbb{F}$.
\item[(b)] A sequence $(h^n)_{n \in \mathbb{N}} \subset L(S,\mathbb{F})$ satisfying that $\int h^n dS$ is $P$-a.s.\ bounded from below, for all $n \in \mathbb{N}$, is said to yield a Free Lunch with Vanishing Risk in $\mathbb{F}$ if there exists $\varepsilon > 0$ and an increasing nonnegative sequence $(\delta_n)_{n \in \mathbb{N}}$ tending to 1 such that $\int_0^T h^n_t dS_t > -1+\delta_n$, $P$-a.s., and $P(\int_0^T h^n_t dS_t > \varepsilon) \geq \varepsilon$, for all $n \in \mathbb{N}$. If there does not exist such a sequence, we say that the {\em No Free Lunch with Vanishing Risk {\bf (NFLVR)}} condition holds for $\mathbb{F}$.
\end{enumerate}
\end{definition}

Now, let us consider that the investor evaluates his preferences by an exogenous time and state independent utility function. 
Intuitively, the utility function $U$ compares the satisfactory of the investor brought by different cash amounts. 
Rigorously, a utility function $U$ is defined in the definition below.

\begin{definition}[Utility function] \label{def utility function}
A function $U:(0,+\infty) \rightarrow \RR \cup \{-\infty\}$, $x \mapsto U(x)$ is called a {\em utility function}, if it is strictly increasing, strictly concave, continuously differentiable and satisfies the \textbf{Inada} conditions:
$$ U'(+\infty):= \lim_{x \rightarrow \infty} U'(x) = 0 \quad \mbox{ and } \quad U'(0) := \lim_{x \searrow 0} U'(x) = +\infty. $$
\end{definition}

{\em Notation:} The inverse function of the first order derivative of $U$ is denoted by $I:=(U')^{-1}$.

\begin{lemma} {\em (cf.\ \cite[(5.15)]{Lakner1995})}
For any $x,y \in (0,+\infty)$ the following inequality holds:
\begin{equation} \label{eq inequality u}
U(I(y)) \ \geq \ U(x) + y \cdot I(y) - xy.
\end{equation}
\end{lemma}

On the other hand, let us consider that the investor's trading is restricted by a risk measure. 
In general, a good risk measure should quantify risk on a monetary scale, detect the risk of extreme loss events and encourage diversification of portfolio choice, as it was pointed out in F\"ollmer \& Schied \cite{Foellmer2002}. 
In this article, we refer to a special risk measure defined through a loss function.

\begin{definition}[Loss function]\label{def loss function}
A function $L:(-\infty,0) \rightarrow \mathbb{R}$ is called {\em loss function}, if it is strictly increasing, strictly convex, continuously differentiable and satisfies $\lim_{x \rightarrow 0} L'(x)>-\infty$ and $\lim_{x \rightarrow -\infty} L'(x)=0$.
\end{definition}

So, we can define a {\em utility-based shortfall risk measure}, as the smallest capital amount $m \in \mathbb{R}$ which has to be added to the position $X$, such that the expected loss function of it stays below some given value $\varepsilon$.

\begin{definition}[Utility-based shortfall risk]\label{def short fall} {\em (cf.\ \cite{Foellmer2002}, p.\ 8-9)}
A risk measure $\rho$ is called {\em utility-based shortfall risk}, if there exists a loss function $L$ defined according to Definition \ref{def loss function}, such that $\rho$ can be written in the form of
$$ \rho(X) \ = \ \inf \left\{ m \in \mathbb{R} \ : \ \mathbb{E}[L(-X - m) \mid \F_0] \, \leq \, \varepsilon \right\}. $$
\end{definition}

A typical example for a utility-based risk measure is the \textit{entropic risk}, defined as
\begin{eqnarray*} \label{def entropic risk}
e_{\gamma}(X) & := & \frac{1}{\gamma} (\ln \mathbb{E} [\exp\{-\gamma X\} \, | \, \F_0] - \ln \varepsilon),
\end{eqnarray*}
where $\gamma > 0$ represents the risk aversion of the investor.

\section{Solution of the optimization problem under different information} \label{sec solution}

\subsection{General optimization problem} \label{subsec general problem}

Let $x \in L^1(P,\F_0)$ denote the positive and exogenously given {\em initial capital} of the investor. 
The set of admissible investment processes which are rigorously defined in the particular subsections is denoted by $\X(\FF,x)$. 
The utility function $U$ and the loss function $L$ as well as the benchmark $\varepsilon \in L^1(P,\F_0)$ are given. 
This paper aims at solving the following portfolio optimization problem under utility-based shortfall risk constraint. 

\begin{problem} \label{problem1}
Find an optimal investment process $\tilde{X}$ that achieves the maximum expected utility under the state of information modeled by the filtration $\mathbb{F}:=(\mathcal{F}_t)_{t \in [0,T]}$
\begin{equation} \label{utility max}
u(x) \ := \ \esssup \limits_{X \in \A(\FF,x)} \mathbb{E} \left[\left. U(X(T)) \, \right| \, \mathcal{F}_0 \right],
\end{equation}
where
\begin{equation*}
\A(\mathbb{F},x) \ := \ \left\{ X \in \X(\FF,x) \, : \, \mathbb{E}[L(-X(T)) \mid \mathcal{F}_0] \leq \varepsilon, \ \mathbb{E} \left[\left. U^-(X(T)) \, \right| \, \mathcal{F}_0 \right] < +\infty, \ P\text{-a.s.} \right\}
\end{equation*}
is the set of admissible investment processes that satisfy the constraint on the utility-based shortfall risk. 
The function $u(\cdot)$ is called the ``value function" of this optimization problem.
\end{problem}

To exclude trivial cases we assume throughout the paper that
\begin{equation} \label{ass u finite}
\begin{split}
& \esssup \limits_{X \in \X(\mathbb{F},x)} \mathbb{E} \left[ U(X(T)) \mid \mathcal{F}_0 \right] \ < \ +\infty, \quad \text{for some } x > 0; \\
& \essinf\limits_{X \in \X(\mathbb{F},x)} \mathbb{E} \left[ L(-X(T)) \mid \mathcal{F}_0\right] \ > \ -\infty, \quad \text{for all } x > 0.
\end{split}
\end{equation}

Possibly, there will not be a solution to this optimization problem for all $\varepsilon$.
On the one hand, the restriction could be too strong that there is no investment process that 
satisfies the risk constraint.
On the other hand, the restriction could also be too weak such that the risk constraint is not binding. To be more precise, let us define
\begin{eqnarray}
\varepsilon_{\min} & := & \essinf_{X \in \X(\FF,x)} \, \left\{ \EE[L(-X(T))|\mathcal{F}_0] \right\}  \qquad \text{and} \label{eq epsilon min} \\
\varepsilon_{\max} & := & \esssup_{X \in \X(\FF,x)} \, \left\{ \EE[L(-X(T))|\F_0] \, : \, \EE[U(X(T))|\mathcal{F}_0] \geq \EE[U(X^*(T))|\F_0] \mbox{ for any } X^* \in \X(\FF,x) \right\}. \label{eq epsilon max}
\end{eqnarray}

From now on, for a given initial capital $x > 0$, utility and loss function $U$, $L$ and filtration $\mathbb{F}$ we assume that these values exist and are finite and we choose $\varepsilon$ such that $\varepsilon_{\min} \leq \varepsilon \leq \varepsilon_{\max}$, $P$-a.s. \\

According to the results of Janke \& Li \cite{Janke2016} for optimization problems under constraints, we reformulate the problem by introducing a Lagrange multiplier $\lambda \geq 0$. Let us define this new function $\tilde{U}_\lambda:(0, +\infty) \to \RR \cup \{-\infty\}$ by
\begin{equation} \label{def W}
\tilde{U}_\lambda(X) \ := \ U(X) - \lambda L(-X), \quad \lambda > 0.
\end{equation}
By the definitions of $U$ and $L$, we have that $\tilde{U}_\lambda$ has the same properties as a usual utility function $U$ defined in Definition \ref{def utility function}: $\tilde{U}_\lambda$ is strictly increasing, strictly concave and continuously differentiable on $(0,+\infty)$. 
Moreover, $\tilde{U}_\lambda$ satisfies the Inada conditions, cf.\ \cite[Proposition 2.11]{Janke2016}.

The inverse function $\tilde{I}_\lambda$ of the first order derivative of $\tilde{U}$ exists and is defined by
\begin{equation}\label{def H}
 \tilde{I}_\lambda \ := \ (\tilde{U}_\lambda')^{-1}.
\end{equation}

\subsection{The initially enlarged filtrations} \label{subsec g w tau}

We introduce two filtrations where the change point is known already at the beginning: 
\begin{itemize}
\item
$\widetilde{\GG}^{W}=(\widetilde{\G}^{W}_t)_{t \geq 0}$, where $\widetilde{\G}^{W}_t:=\bigcap_{s>t} (\mathcal{F}^W_s \vee \sigma(\tau))$ represents the knowledge of the Brownian motion $W$ up to time $t$ plus the knowledge of the random time $\tau$ already at time $t=0$,
\item
$\widetilde{\GG}^{S}=(\widetilde{\G}^{S}_t)_{t \geq 0}$, where $\widetilde{\G}^{S}_t:=\bigcap_{s>t} (\mathcal{F}^S_s \vee \sigma(\tau))$ represents the knowledge of the price process $S$ of the risky asset up to time $t$ plus the knowledge of the random time $\tau$ already at time $t=0$. 
\end{itemize}

These filtrations are motivated by financial markets where an investor has insider information, e.\,g.\ he knows the time of an interest rate change. \\

Since it holds that $\widetilde{\GG}^{S} \subseteq \widetilde{\GG}^{W}$ and of course $\widetilde{\G}^{W}_0=\widetilde{\G}^{S}_0$, the results for these two filtrations are very similar. 
For dealing with the optimization problem defined in Problem \ref{problem1} we need the following assumptions.

\begin{assumption} 
$ $ \label{ass 3}
\begin{enumerate} 
\item[(a)] 
There exists a $\widetilde{\GG}^{W}$-predictable process $\tilde{\theta}=(\tilde{\theta}_t)_{t \in [0,T]}$ and a $\widetilde{\GG}^{W}$-Brownian motion $\widetilde{W}=(\widetilde{W}_t)_{t \in [0,T]}$ such that
$$ W_t \ = \ \widetilde{W}_t + \int_0^t \tilde{\theta}_u \, du, \qquad \mbox{for all } t \in [0,T]. $$ 
Moreover, let us define the $\widetilde{\GG}^{W}$-predictable process $\tilde{\mu}^{W}:=(\tilde{\mu}^{W}_{t})_{t \in [0,T]}$ by
\begin{eqnarray} 
\tilde{\mu}^{W}_{t} & := & {\bf 1}_{\{t \leq \tau\}} \left(\mu^{1}(t,\tilde{S}_{t}) + \sigma^{1}(t,\tilde{S}_{t})\tilde{\theta}_{t} \right) + {\bf 1}_{\{t > \tau\}} \left(\mu^{2}(t,\tilde{S}_{t}) + \sigma^{2}(t,\tilde{S}_{t}) \tilde{\theta}_{t} \right). \label{eq mu for g tau} 
\end{eqnarray} 
\item[(b)]
Every $\widetilde{\GG}^{W}$-local martingale $\widetilde{L}^W=(\widetilde{L}^W_t)_{t \in [0,T]}$ admits the representation
$$ \widetilde{L}^W_t \ := \ \widetilde{L}^W_{0} + \int_{0}^{t} \widetilde{\varphi}^{W}_{u} \, d\widetilde{W}^{(\tau)}_{u}, \qquad 0 \leq t \leq T, $$
for some $\widetilde{\GG}^{W}$-predictable process $\widetilde{\varphi}^{W}=(\widetilde{\varphi}^{W}_{t})_{t \in [0,T]}$ satisfying $\int_{0}^{T} (\widetilde{\varphi}^{W}_{t})^{2} dt < +\infty$, $P$-a.s. 
\item[(c)] 
It holds (NFLVR) in the filtration $\widetilde{\mathbb{G}}^{W}$ and the process $\widetilde{Z}^{W}=(\widetilde{Z}^{W}_t)_{t \in [0,T]}$ defined by $\widetilde{Z}^{W}_t := \E(\int \frac{\tilde{\mu}^{W}}{\sigma} d\widetilde{W})_t$
is a true martingale. 
Moreover, there is a unique equivalent probability measure $\widetilde{P}^W \sim P$ such that $\frac{d\widetilde{P}^W}{dP} = \widetilde{Z}^{W}_T$, which is equivalent to market completeness. 
\end{enumerate}
\end{assumption}

For a better understanding of these assumptions, let us give some further explanations. 

\begin{remark} 
$ $ \label{remark condition g w tau}
\begin{enumerate}
\item[(a)] 
Statements (a) and (b) hold true if the {\em density hypothesis} holds, i.\,e.\ the $\F_t^W$-conditional law of the random time $\tau$ admits a density w.\,r.\,t.\ to the unconditional law of $\tau$, for all $t \in [0,T]$, cf.\ \cite{jacod}. 
This is equivalent to the existence of a probability measure $R \sim P$ such that $\G_T^W$\footnote{For the definition of the $\sigma$-algebra $\G_T^W$ see Subsection \ref{subsec g w}.} and $\widetilde{\G}^{W}_0$ are $R$-independent. This implies the existence of a unique martingale preserving probability measure, cf.\ \cite[Chapter 2]{Amendinger2003}.
Although the density hypothesis is not valid if $\tau$ is a stopping time w.\,r.\,t.\ to the filtration $\FF^W$, one can use results by \cite[Chapter 12]{Yor1997} and \cite[Hypothesis 3]{Jeanblanc2015a} to satisfy statement (a).
\item[(b)] 
If (NFLVR) holds true, then, by \cite[Prop.\ 5.2.]{Fontana2013}, it holds that $\int_{0}^{T} (\tilde{\theta}_{u})^{2}<+\infty$, $P$-a.s., which is equivalent to $\int_{0}^{T}(\tilde{\mu}^{W}_{u}/\sigma_{u})^{2} du < +\infty$, $P$-a.s.
Moreover, it holds for the process $\widetilde{Z}^{W}=(\widetilde{Z}^{W}_t)_{t \in [0,T]}$ that $\widetilde{Z}^{W}S \in \mathcal{M}_{loc}(\widetilde{\GG}^{W})$, or, equivalently, $\widetilde{Z}^{W}\tilde{S} \in \mathcal{M}_{loc}(\widetilde{\GG}^{W})$. 
If $\tilde{\mu}^{W}/\sigma$ is uniformly bounded, then $\widetilde{Z}^{W}$ is indeed a true martingale, cf.\ \cite[p.\ 250]{Lakner1995}.
\item[(c)]
Without statement (c) we cannot hope to have a complete market for this filtration. 
Since the $\sigma$-algebra $\widetilde{\G}^{W}_0$ is not trivial, one cannot find a unique measure $\widetilde{P} \sim P$ 
with $S \in \M_{loc}(\widetilde{P},\widetilde{\GG}^{W})$ because there is no condition in the choice of $\widetilde{P}$ on the field $\widetilde{\G}^{W}_0$.
\end{enumerate}
\end{remark}

Under Assumptions \ref{ass 1} and \ref{ass 3} (a)  the process $\tilde{S}=(\tilde{S}_{t})_{t \in [0,T]}$ has the following representation w.\,r.\,t.\ the filtration $\widetilde{\GG}^{W}$:
\begin{equation} \label{eq repr x for g tau}
\tilde{S}_{t} \ = \ \int_{0}^{t} \tilde{\mu}^{W}_{u} \, du + \int_{0}^{t} \sigma_{u} \, d\widetilde{W}_{u}, \qquad \mbox{for all } t \in [0,T].
\end{equation}

Let us now consider the price process representation as well as the martingale representation w.\,r.\,t.\ filtration $\widetilde{\GG}^{S}$. 

\begin{lemma} {\em (cf.\ \cite[Lemma 5.5., Proposition 5.8.]{Fontana2013})} \label{lemma representation g s tau}
Suppose Assumptions \ref{ass 1} and \ref{ass 3} (a), (c) hold true. 
\begin{enumerate}
\item[(i)] 
The process $\tilde{S}=(\tilde{S}_{t})_{t \in [0,T]}$ has the following representation w.\,r.\,t.\ the filtration $\widetilde{\GG}^{S}$:
\begin{equation} \label{eq repr x for g s tau}
\tilde{S}_{t} \ = \ \int_{0}^{t} \tilde{\mu}^{S}_{u} \, du + \int_{0}^{t} \sigma_{u} \, d\widetilde{W}^{S}_{u}, \qquad \mbox{for all } t \in [0,T],
\end{equation}
where the $\widetilde{\GG}^{S}$-predictable process $\tilde{\mu}^{S}:=(\tilde{\mu}^{S}_{t})_{t \in [0,T]}$ is defined by
\begin{eqnarray} 
\tilde{\mu}^{S}_{t} & := & {\bf 1}_{\{t \leq \tau\}} \left(\mu^{1}(t,\tilde{S}_{t}) + \sigma^{1}(t,\tilde{S}_{t}) \, ^{p}\tilde{\theta}_{t} \right) + {\bf 1}_{\{t > \tau\}} \left(\mu^{2}(t,\tilde{S}_{t}) + \sigma^{2}(t,\tilde{S}_{t}) \,  ^{p}\tilde{\theta}_{t} \right), \label{eq mu for g s tau}
\end{eqnarray}
where $^{p}\tilde{\theta}$ denotes the $\widetilde{\GG}^{S}$-predictable projection\footnote{For the exact definition and properties we refer the reader to (cf.\ \cite[p.\ 358]{nikeghbali}).} of $\tilde{\theta}$, the process $\widetilde{W}^{S}=(\widetilde{W}^{S}_{t})_{t \in [0,T]}$ is a $\widetilde{\GG}^{W}$-Brownian motion independent of $\tau$, and $\sigma$ is defined as in \eqref{eq sigma general}. 
Note that $\tilde{\mu}^S=\,^p\tilde{\mu}^W$. 
\item[(ii)]
Every $\widetilde{\GG}^{S}$-local martingale $\widetilde{L}^{S}=(\widetilde{L}^{S}_t)_{t \in [0,T]}$ has the form
$$ \widetilde{L}^{S}_t \ := \ \widetilde{L}^{S}_{0} + \int_{0}^{t} \widetilde{\varphi}^{S}_{u} \, d\widetilde{W}^S_{u}, \qquad 0 \leq t \leq T, $$
for some $\widetilde{\GG}^{S}$-predictable process $\widetilde{\varphi}^{S}=(\widetilde{\varphi}^{S}_{t})_{t \in [0,T]}$ satisfying $\int_{0}^{T} (\widetilde{\varphi}^{S}_{t})^{2} dt < +\infty$, $P$-a.s.
\end{enumerate}
\end{lemma}

By Assumption \ref{ass 3} (c), (NFLVR) also holds in the smaller filtration $\widetilde{\GG}^{S}$, so the $\widetilde{\GG}^{S}$-adapted process $\widetilde{Z}^{S}=(\widetilde{Z}^{S}_{t})_{t \in [0,T]}$ defined by $\widetilde{Z}^{S}_t=\E(\int \frac{\tilde{\mu}^{S}}{\sigma} d\widetilde{W}^{S})_t$ is a true martingale and we can define $\widetilde{P}^S \sim P$ as a probability measure such that $ \frac{d\widetilde{P}^S}{dP} = \widetilde{Z}^{S}_T$. \\
From now on if we state results holding for both filtrations, we simply write $\bullet \in \{W,S\}$, so e.\,g.\ $\widetilde{\GG}^{\bullet}$ for $\widetilde{\GG}^{W}$ and $\widetilde{\GG}^{S}$ or $\widetilde{\G}^{\bullet}_t$ for $\widetilde{\G}^{W}_t$ and $\widetilde{\G}^{S}_t$, $t \in [0,T]$, respectively. 
Note, that $\widetilde{W}^W:=\widetilde{W}$. \\
Moreover, by $\widetilde{\EE}^\bullet$ we denote the expectation w.\,r.\,t.\ $\widetilde{P}^\bullet$. \\

Let us now define portfolio processes for this subsection. 
We assume that it is \textit{self-financing}, i.\,e.\ there will be no exogenous cash-flow like credits or consumption. 
A portfolio process represents the amount of money held in the risky asset. 

\begin{definition} \label{def trading strategy g w tau}
A process $\pi=(\pi_t)_{t \in [0,T]} \in L(\widetilde{\GG}^{\bullet},S)$ satisfying $\int_{0}^{T} (\pi_{u}\sigma_{u})^{2} du < +\infty$, $P$-a.s., is called {\em trading strategy} and the corresponding investment process is denoted by
\begin{equation} \label{eq wealth process v in g tau}
X^{\pi}(t) \ = \ x + \int_{0}^{t} \pi_{u} \, d\tilde{S}_{u} \ = \ x + \int_{0}^{t} \pi_{u}\tilde{\mu}^{\bullet}_{u}\, du + \int_{0}^{t} \pi_{u} \sigma_{u} \, d\widetilde{W}^{\bullet}_{u}, \qquad 0 \leq t \leq T,
\end{equation}
where the initial capital $x \in L^1(P,\widetilde{\G}^{\bullet}_0)$ is assumed to be greater than zero. 
$\pi$ is called {\em admissible} if the corresponding wealth process $X^\pi(t)$ is non-negative for all $t \in [0,T]$, $P$-a.s. 
The sets of admissible trading strategies and of the corresponding investment processes are denoted by $\Pi(\widetilde{\GG}^{\bullet},x)$ and $\X(\widetilde{\GG}^{\bullet},x)$, respectively. 
\end{definition}

{\em Remark.} 
By the assumption of (NFLVR), equation \eqref{eq wealth process v in g tau} is well-defined, since by the Cauchy-Schwarz inequality it holds:
$$ \int_0^T \pi_t \tilde{\mu}^{\bullet}_t \, dt \ = \ \int_0^T \pi_t \sigma_t \frac{\tilde{\mu}^{\bullet}_t}{\sigma_t} \, dt \ \leq \ \left(\int_0^T (\pi_t \sigma_t)^2 \, dt\right)^\frac12 \cdot \left( \int_0^T \left(\frac{\tilde{\mu}^{\bullet}_t}{\sigma_t}\right)^2 \, dt\right)^\frac12 \ < \ + \infty, \ P\mbox{-a.s.} $$ 

Define the process $\widetilde{B}^{\bullet}=(\widetilde{B}^{\bullet}_{t})_{t \in [0,T]}$ by 
$$ \widetilde{B}^{\bullet}_{t} \ = \ \widetilde{W}^{\bullet}_{t} + \int_{0}^{t} \frac{\tilde{\mu}^{\bullet}_{u}}{\sigma_{u}} \, du, \qquad 0 \leq t \leq T, $$
which is a Brownian motion under $\widetilde{P}^\bullet$ w.\,r.\,t.\ the filtration $\widetilde{\GG}^{\bullet}$. 
So, the process $\tilde{S}$ is of the form
\begin{eqnarray*}
\tilde{S}_{t} & = & \int_{0}^{t} \tilde{\mu}^{\bullet}_{u} \, du + \int_{0}^{t} \sigma_{u} \, d\widetilde{W}^{\bullet}_{u} \ = \ \int_{0}^{t} \sigma_{u} \, d\widetilde{B}^{\bullet}_{u}, \qquad  \mbox{for all } t \in [0,T],
\end{eqnarray*}
and for $\pi \in \Pi(\widetilde{\GG}^{\bullet},x)$ the wealth process $X^{\pi}$ admits the representation
\begin{eqnarray*}
X^{\pi}(t) & = & x + \int_{0}^{t} \pi_{u} \sigma_{u} \, d\widetilde{B}^{\bullet}_{u}, \qquad  \mbox{for all } t \in [0,T].
\end{eqnarray*}

We have the following main result:

\begin{theorem} \label{theorem optimal trading strategy g w tau}
Suppose Assumptions \ref{ass 1} and \ref{ass 3} hold true. 
Moreover, let $\EE[\widetilde{Z}^{\bullet}_T \cdot \tilde{I}_\lambda(y\widetilde{Z}^{\bullet}_T) | \widetilde{\G}^{\bullet}_{0}]<+\infty$ for all $\widetilde{\G}^{\bullet}_0$-measurable $y \in (0,+\infty)$, $P$-a.s. 
Then the optimal terminal wealth $\hat{R}$ for Problem \ref{problem1} is given by
$$ \hat{R} \ = \ \tilde{I}_{\lambda^*}\left( \hat{y}\widetilde{Z}^{\bullet}_T \right), $$
where the $\widetilde{\G}^{\bullet}_0$-measurable random variables $\hat{y} \in (0,+\infty)$ and $\lambda^* \geq 0$ are such that
\begin{equation} \label{eq lambda y} 
\left\{ \begin{array} {rcl}  \EE\left[\widetilde{Z}^{\bullet}_T \cdot \tilde{I}_{\lambda^*}(\hat{y}\widetilde{Z}^{\bullet}_T) \mid \widetilde{\G}^{\bullet}_{0}\right] & =& x, \quad P\text{-a.s.,} \\ 
\EE\left[L(-\tilde{I}_{\lambda^*}(\hat{y}\widetilde{Z}^{\bullet}_T)) \mid \widetilde{\G}^{\bullet}_0 \right] & = & \varepsilon, \quad P\text{-a.s.}. \end{array} \right.
\end{equation}
The unique optimal trading strategy $\hat{\pi} \in \Pi(\widetilde{\GG}^{\bullet},x)$ and the corresponding investment process $X^{\hat{\pi}}$ satisfy
$$ X^{\hat{\pi}}(t) \ = \ x + \int_{0}^{t} \hat{\pi}_{u} \sigma_{u} \, d\widetilde{B}^{\bullet}_{u} \ = \ \widetilde{\EE}^\bullet\left[ \hat{R} \, \left| \, \widetilde{\G}^{\bullet}_{t} \right.\right], \qquad 0 \leq t \leq T. $$
\end{theorem}

To prove this theorem, we first formulate a martingale representation theorem for the measure $\widetilde{P}^\bullet$ w.\,r.\,t.\ the filtration $\widetilde{\GG}^{\bullet}$.

\begin{lemma} \label{mrt g w tau}
Suppose Assumptions \ref{ass 1} and \ref{ass 3} hold true. 
Every local martingale $\widetilde{M}^\bullet=(\widetilde{M}^\bullet_{t})_{t \in [0,T]}$ under $\widetilde{P}^\bullet$ w.\,r.\,t.\ the filtration $\widetilde{\GG}^{\bullet}$ is of the form
\begin{equation} \label{eq mrt for tilde p in g tau 2} 
\widetilde{M}^\bullet_{t} \ = \ \widetilde{M}^\bullet_{0} + \int_{0}^{t} \pi_{u} \sigma_{u} \, d\widetilde{B}^{\bullet}_{u}, \qquad 0 \leq t \leq T,
\end{equation}
for some $\widetilde{\GG}^{\bullet}$-predictable process $\pi=(\pi_t)_{t \in [0,T]}$ satisfying $\int_{0}^{T} (\pi_{u}\sigma_{u})^{2} du < +\infty$, $P$-a.s. 
\end{lemma}

{\em Proof.}
First, let us show that the process $\widetilde{Z}^{\bullet}\widetilde{M}^\bullet=(\widetilde{Z}^{\bullet}_t \widetilde{M}^\bullet_t)_{t \in [0,T]}$ is a local $P$-martingale. 
Since $\widetilde{M}^\bullet$ is a local $\widetilde{P}^\bullet$-martingale, there exists an increasing sequence of stopping times $(T_n)_{n \in \NN}$ satisfying $T_n \to +\infty$. 
Now it holds for any $0 \leq s \leq t \leq T$ and $n \in \NN$ by Bayes' formula:
$$ \EE\left[\widetilde{Z}^{\bullet}_{t \wedge T_n} \widetilde{M}^\bullet_{t \wedge T_n} \, | \, \widetilde{\G}^{\bullet}_{s}\right] \ = \ \widetilde{Z}^{\bullet}_{s \wedge T_n} \widetilde{\EE}^\bullet\left[ \widetilde{M}^\bullet_{t \wedge T_n} \, | \, \widetilde{\G}^{\bullet}_{s}\right] \ = \ \widetilde{Z}^{\bullet}_{s \wedge T_n} \widetilde{M}^\bullet_{s \wedge T_n}. $$
By the martingale representation property (Assumption \ref{ass 3} (b) and Lemma \ref{lemma representation g s tau} (b)), the process $\widetilde{L}^{\bullet}:=\widetilde{Z}^{\bullet}\widetilde{M}^\bullet$ has the form
$$ \widetilde{L}^{\bullet}_t \ = \ \widetilde{L}^\bullet_{0} + \int_{0}^{t} \widetilde{\varphi}^{\bullet}_{u} \, d\widetilde{W}^{\bullet}_{u}, \qquad 0 \leq t \leq T, $$
for some $\widetilde{\GG}^{\bullet}$-predictable process $\widetilde{\varphi}^\bullet=(\widetilde{\varphi}^\bullet_{t})_{t \in [0,T]}$ satisfying $\int_{0}^{T} (\widetilde{\varphi}^\bullet_{t})^{2} dt < +\infty$, $P$-a.s. 
Then, we have $\widetilde{M}^\bullet=\widetilde{L}^\bullet (\widetilde{Z}^{\bullet})^{-1}$, where the dynamics of $(\widetilde{Z}^{\bullet})^{-1}$ by It\^{o}'s rule is
\begin{eqnarray*}
d(\widetilde{Z}^{\bullet})^{-1}_t & = & -(\widetilde{Z}^{\bullet})^{-1}_t \, d\left(-\int_{0}^{t} \frac{\tilde{\mu}^{\bullet}_{u}}{\sigma_{u}} \, d\widetilde{W}^{\bullet}_{u} - \frac12 \int_{0}^{t} \left(\frac{\tilde{\mu}^{\bullet}_{u}}{\sigma_{u}}\right)^{2} \, du \right) + \frac12 (\widetilde{Z}^{\bullet})^{-1}_t \left(\frac{\tilde{\mu}^{\bullet}_{t}}{\sigma_{t}}\right)^{2} \, dt \\
& = & \frac{\tilde{\mu}^{\bullet}_{t}}{\sigma_{t}} (\widetilde{Z}^{\bullet})^{-1}_t \, d\widetilde{W}^{\bullet}_t + \left(\frac{\tilde{\mu}^{\bullet}_{t}}{\sigma_{t}}\right)^{2}(\widetilde{Z}^{\bullet})^{-1}_t \, dt.
\end{eqnarray*}
Using It\^{o}'s product rule, the dynamics of $\widetilde{M}^{\bullet}$ is given by:
\begin{eqnarray*}
d\widetilde{M}^{\bullet}_t & = & d\left(\widetilde{L}^{\bullet} (\widetilde{Z}^{\bullet})^{-1}\right)_t \\
& = & \widetilde{L}^{\bullet}_t \, d(\widetilde{Z}^{\bullet})^{-1}_t + (\widetilde{Z}^{\bullet})^{-1}_t \, d\widetilde{L}^{\bullet}_t + d \langle \widetilde{L}^{\bullet}, (\widetilde{Z}^{\bullet})^{-1} \rangle_t \\
& = & \widetilde{Z}^{\bullet}_{t}\widetilde{M}^{\bullet}_{t} \cdot \left(\frac{\tilde{\mu}_{t}^{\bullet}}{\sigma_{t}} (\widetilde{Z}^{\bullet})^{-1}_t \, d\widetilde{W}^{\bullet}_t + \left(\frac{\tilde{\mu}_{t}^{\bullet}}{\sigma_{t}}\right)^{2}(\widetilde{Z}^{\bullet})^{-1}_t \, dt \right) + (\widetilde{Z}^{\bullet})^{-1}_t \widetilde{\varphi}^{\bullet}_t \, d\widetilde{W}^{\bullet}_t + (\widetilde{Z}^{\bullet})^{-1}_t \widetilde{\varphi}^{\bullet}_t \frac{\tilde{\mu}^{\bullet}_{t}}{\sigma_{t}} \, dt \\
& = & \widetilde{Z}^{\bullet}_{t}\widetilde{M}^{\bullet}_{t} \cdot \frac{\tilde{\mu}^{\bullet}_{t}}{\sigma_{t}} (\widetilde{Z}^{\bullet})^{-1}_t \, d\widetilde{W}^{\bullet}_t + \widetilde{Z}^{\bullet}_{t}\widetilde{M}^{\bullet}_{t} \cdot \left(\frac{\tilde{\mu}^{\bullet}_{t}}{\sigma_{t}}\right)^{2}(\widetilde{Z}^{\bullet})^{-1}_t \, dt + (\widetilde{Z}^{\bullet})^{-1}_t \widetilde{\varphi}^{\bullet}_t \, d\widetilde{W}^{\bullet}_t + (\widetilde{Z}^{\bullet})^{-1}_t \widetilde{\varphi}^{\bullet}_t \frac{\tilde{\mu}^{\bullet}_{t}}{\sigma_{t}} \, dt \\
& = & \left(\widetilde{M}^{\bullet}_{t} \frac{\tilde{\mu}^{\bullet}_{t}}{\sigma_{t}} + (\widetilde{Z}^{\bullet})^{-1}_t \widetilde{\varphi}^{\bullet}_t\right) \, \left(d\widetilde{W}^{\bullet}_t + \frac{\tilde{\mu}^{\bullet}_{t}}{\sigma_{t}} \, dt \right) \\
& = & \left(\widetilde{M}^{\bullet}_{t} \frac{\tilde{\mu}^{\bullet}_{t}}{\sigma_{t}} + (\widetilde{Z}^{\bullet})^{-1}_t \widetilde{\varphi}^{\bullet}_t\right) \, d\widetilde{B}^{\bullet}_{t}.
\end{eqnarray*}
Set $\tilde{\phi}^{\bullet}_{t}:=\widetilde{M}^{\bullet}_{t} \frac{\tilde{\mu}^{\bullet}_{t}}{\sigma_{t}} + (\widetilde{Z}^{\bullet})^{-1}_t \widetilde{\varphi}^{\bullet}_t$, then $\widetilde{M}^{\bullet}$ is of the form
\begin{equation*} 
\widetilde{M}^{\bullet}_{t} \ = \ \widetilde{M}^{\bullet}_{0} + \int_{0}^{t} \tilde{\phi}^{\bullet}_{u} \, d\widetilde{B}^{\bullet}_{u}, \qquad 0 \leq t \leq T, 
\end{equation*}
where $\tilde{\phi}^{\bullet}=(\tilde{\phi}^{\bullet}_{t})_{t \in [0,T]}$ is a $\widetilde{\GG}^{\bullet}$-predictable process satisfying $\int_{0}^{T} (\tilde{\phi}^{\bullet}_{t})^{2} dt < +\infty$, $P$-a.s., which follows from Remark \ref{remark condition g w tau}(b) and the path continuity of $\widetilde{Z}^{\bullet}$ and $\widetilde{M}^{\bullet}$, cf.\ \cite{Karatzas1991}. 
Set $\tilde{\phi}^{\bullet}= \pi \sigma$ for some $\widetilde{\GG}^{\bullet}$-predictable process $\pi$, so the claim follows.
\hfill{$\Box$} \\

With this result we are now able to give the proof of the main result. \\

{\bf Proof of Theorem \ref{theorem optimal trading strategy g w tau}.}
The existence of $\hat{y} \in (0,+\infty)$ and $\lambda^* \geq 0$ such that equations \eqref{eq lambda y} hold is proven in Appendix \ref{app lagrangian}. 
Next we show that there exists a trading strategy $\hat{\pi} \in \Pi(\widetilde{\GG}^{\bullet},x)$ such that for the corresponding wealth process $X^{\hat{\pi}}$ it holds:
$$ X^{\hat{\pi}}(t) \ = \ \widetilde{\EE}^\bullet\left[ \hat{R} \, \left| \, \widetilde{\G}^{\bullet}_{t} \right.\right]. $$
Define $M^\bullet(t) = \widetilde{\EE}^\bullet\left[\hat{R} \, | \, \widetilde{\G}^{\bullet}_{t}\right]$, $0 \leq t \leq T$. 
Because for $0 \leq s \leq t$ it holds:
\begin{eqnarray*}
\widetilde{\EE}^\bullet\left[M^\bullet(t) \, \left| \, \widetilde{\G}^{\bullet}_{s} \right.\right] & = & \widetilde{\EE}^\bullet\left[\left. \widetilde{\EE}^\bullet\left[\hat{R} \, \left| \, \widetilde{\G}^{\bullet}_{t}\right.\right] \, \right| \, \widetilde{\G}^{\bullet}_{s}\right] 
\ = \ \widetilde{\EE}^\bullet\left[\hat{R} \, \left| \, \widetilde{\G}^{\bullet}_{s} \right.\right] 
\ = \ M^\bullet(s),
\end{eqnarray*}
$M^\bullet$ is a $\widetilde{P}^\bullet$-martingale w.\,r.\,t.\ the filtration $\widetilde{\GG}^{\bullet}$. 
By Lemma \ref{mrt g w tau}, we obtain that 
$$ M^\bullet(t) \ = \ M^\bullet(0) + \int_{0}^{t} \hat{\pi}_{u} \sigma_{u} \, d\widetilde{B}^{\bullet}_{u}, \qquad 0 \leq t \leq T, $$
with $M^\bullet(0)=\widetilde{\EE}^\bullet[\hat{R} \, | \, \widetilde{\G}^{\bullet}_{0}] = x$ for some $\widetilde{\GG}^{\bullet}$-predictable process $\hat{\pi}$ satisfying $\int_{0}^{T} (\hat{\pi}_{u}\sigma_{u})^{2} du < +\infty$, $P$-a.s. 
So, it holds that $X^{\hat{\pi}}(t) = M^\bullet(t) = \widetilde{\EE}^\bullet[\hat{R} \mid \widetilde{\G}^{\bullet}_t] \geq 0$, since $\hat{R}$ is non-negative. 
Therefore, $\hat{\pi}$ is indeed in $\Pi(\widetilde{\GG}^{\bullet},x)$. \\
Now, let us show that $\EE[U^-(X^{\hat{\pi}}(T)) | \widetilde{\G}^{\bullet}_{0}] < +\infty$. 
By inequality \eqref{eq inequality u} we have
$$ U^-(X^{\hat{\pi}}(T)) \ \leq \ U^-(K) - \hat{y} \widetilde{Z}^{\bullet}_T K $$
for some $K > 0$ such that $U^-(K)<+\infty$. 
By taking expectations and by $\EE[\widetilde{Z}^{\bullet}_T \mid \widetilde{\G}^{\bullet}_{0}]=1$, the claim follows and $X^{\hat{\pi}}$ is indeed in $\A(\widetilde{\GG}^{\bullet},x)$. \\
For the optimality of $X^{\hat{\pi}}$, let us consider another $\pi \in \Pi(\widetilde{\GG}^{\bullet},x)$. 
It holds that $\widetilde{\EE}^\bullet[X^\pi(T) \mid \widetilde{\G}^{\bullet}_{0}] \leq x$, since $\widetilde{P}^\bullet \in \Q^e$ and $X^\pi$ is non-negative. 
We consider the unconstrained optimization problem 
$$ \max_{X \in \X(\widetilde{\GG}^{\bullet},x)} \ \EE \left[ \tilde{U}_{\lambda^*} \left(X(T)\right) \, \left| \, \widetilde{\G}^{\bullet}_{0} \right.\right], $$
where $\tilde{U}_{\lambda^*}$ is a combined utility and loss function as defined in \eqref{def W} and by applying inequality \eqref{eq inequality u} we have:
\begin{eqnarray*}
\tilde{U}_{\lambda^*}(\hat{R}) \ = \ \tilde{U}_{\lambda^*}\left(\tilde{I}(\hat{y}\widetilde{Z}^{\bullet}_T)\right) 
\ = \ \tilde{U}_{\lambda^*}\left(X^{\hat{\pi}}(T)\right) & \geq & \tilde{U}_{\lambda^*}\left(X^\pi(T)\right) + \hat{y}\widetilde{Z}^{\bullet}_T \cdot \hat{R} - \hat{y}\widetilde{Z}^{\bullet}_T \cdot X^\pi(T).
\end{eqnarray*}
Taking conditional expectations on both sides, it follows:
\begin{eqnarray*}
\lefteqn{\EE\left[\tilde{U}_{\lambda^*}(X^{\hat{\pi}}(T)) \, \left| \, \widetilde{\G}^{\bullet}_{0} \right.\right] \ \geq \ \EE \left[\tilde{U}_{\lambda^*}(X^\pi(T)) + \hat{y}\widetilde{Z}^{\bullet}_T \cdot \hat{R} - \hat{y}\widetilde{Z}^{\bullet}_T \cdot X^\pi(T) \, \left| \, \widetilde{\G}^{\bullet}_{0} \right.\right]} \\
 & & = \EE \left[\tilde{U}_{\lambda^*}(X^\pi(T)) \, \left| \, \widetilde{\G}^{\bullet}_{0} \right.\right] + \hat{y} \left( \EE\left[\widetilde{Z}^{\bullet}_T \cdot \hat{R} \, \left| \, \widetilde{\G}^{\bullet}_{0} \right.\right] - \EE\left[\widetilde{Z}^{\bullet}_T \cdot X^\pi(T) \, | \, \widetilde{\G}^{\bullet}_{0} \right] \right) \\
 & & = \EE \left[\tilde{U}_{\lambda^*}(X^\pi(T)) \, \left| \, \widetilde{\G}^{\bullet}_{0} \right.\right] + \hat{y}\left( \underbrace{\widetilde{\EE}^\bullet\left[\hat{R} \, \left| \, \widetilde{\G}^{\bullet}_{0} \right.\right]}_{= \ x} - \underbrace{\widetilde{\EE}^\bullet\left[ X^\pi(T) \, \left| \, \widetilde{\G}^{\bullet}_{0} \right.\right]}_{\leq \ x}\right) \ \geq \ \EE \left[\tilde{U}_{\lambda^*}(X^\pi(T)) \, \left| \, \widetilde{\G}^{\bullet}_{0} \right.\right].
\end{eqnarray*}
Since $X^{\hat{\pi}}$ satisfies the risk constraint as well as $\EE[U^-(X^{\hat{\pi}}) \mid \widetilde{\G}^{\bullet}_{0}]$ it is in $\A(\widetilde{\GG}^{\bullet},x)$ and therefore an optimal solution for Problem \ref{problem1}. \\
Last, we will show the uniqueness of $\hat{\pi}$. 
This will be done in several steps: First, let us show that the optimal investment process $X^{\hat{\pi}}$ is unique: 
For that, let us suppose there are $\pi^1,\pi^2 \in \Pi(\widetilde{\GG}^{\bullet},x)$ such that $X^{\pi^1}, X^{\pi^2}$ are optimal for Problem \ref{problem1}.
Then we define $\pi^3$ such that $X^{\pi^3}=\frac12(X^{\pi^1} + X^{\pi^2})$. 
By the convexity of the loss function $L$ and the function $U^-$ we have that $X^{\pi^3} \in \X(\widetilde{\GG}^{\bullet},x)$. 
But on the other hand it holds due to the concavity of $U$ that
\begin{eqnarray*}
\EE \left[U(X^{\pi^3}(T)) \mid \widetilde{\G}^{\bullet}_0 \right] & = & \EE \left[U(1/2(X^{\pi^1}(T)+X^{\pi^2}(T))) \mid \widetilde{\G}^{\bullet}_0 \right] \\ 
& \geq & \frac12 \EE \left[U(X^{\pi^1}(T))+U(X^{\pi^2}(T))) \mid \widetilde{\G}^{\bullet}_0 \right] \ = \ \EE \left[U(X^{\pi^1}(T)) \mid \widetilde{\G}^{\bullet}_0 \right],
\end{eqnarray*}
where we have equality in the second line due to the optimality of $X^{\pi^1}$ and $X^{\pi^2}$. 
Therefore, we must have that $X^{\pi^1}(T)=X^{\pi^2}(T)$, $P$-a.s. \\
Next, we show that the optimal investment process has to satisfy \eqref{eq lambda y}.
Let us start with $\widetilde{\EE}^\bullet[X^{\hat{\pi}}(T) \mid \widetilde{\G}^{\bullet}_0 ] = x$, $P$-a.s. 
First suppose that there exists a subset $\Sigma \subset \Omega$ satisfying $P(\Sigma) >0$ such that $\widetilde{\EE}^\bullet[X^{\hat{\pi}}(T) \mid \widetilde{\G}^{\bullet}_0](\omega) = 0$ for all $\omega \in \Sigma$. 
Obviously we have that $X^{\hat{\pi}}(T)\equiv 0$ on $\Sigma$, since $\hat{\pi} \in \Pi(\GG^{\bullet,(\tau)},x)$. 
Let us consider the trading strategy $\tilde{\pi} \equiv 0$. 
It holds that $X^{\tilde{\pi}}(t)=x$, $t \in [0,T]$. 
Especially we have that $X^{\tilde{\pi}}(T) > X^{\hat{\pi}}(T)$, $P$-a.s., and by concavity of $U$ and convexity of $L$ and $U^-$ it holds that $X^{\tilde{\pi}} \in \X(\widetilde{\GG}^{\bullet},x)$ and 
$$ \EE \left[U(X^{\tilde{\pi}}(T)) \mid \widetilde{\G}^{\bullet}_0 \right] \ > \ \EE \left[U(X^{\hat{\pi}}(T)) \mid \widetilde{\G}^{\bullet}_0 \right], \qquad P\text{-a.s.,} $$
which contradicts the optimality of $X^{\hat{\pi}}$. \\
Now, suppose there is another subset $\Sigma^* \subset \Omega$ with $P(\Sigma^*)>0$ such that $\widetilde{\EE}^\bullet[X^{\hat{\pi}}(T) \mid \widetilde{\G}^{\bullet}_0](\omega) = a_\omega < x(\omega)$ for all $\omega \in \Sigma^*$. 
Choose a trading strategy $\pi^*$ such that $X^{\pi^*}(\omega)=\frac{x(\omega)}{a_\omega}X^{\hat{\pi}}(\omega)$. 
It holds that $\widetilde{\EE}^\bullet[X^{\pi^*}(T) \mid \widetilde{\G}^{\bullet}_0](\omega) =x(\omega)$ and $X^{\pi^*}(\omega)>X^{\hat{\pi}}(\omega)$. 
By the monotonicity of $U$, $L$ and $U^-$ we have that $X^{\pi^*} \in \X(\widetilde{\GG}^{\bullet},x)$ and $\EE[U(X^{\pi^*}(T) \mid \widetilde{\G}^{\bullet}_0] > \EE[U(X^{\hat{\pi}}(T)) \mid \widetilde{\G}^{\bullet}_0]$, $P$-a.s.; a contradiction to the optimality of $X^{\hat{\pi}}$. \\
Last, the condition $\EE[L(-X^{\hat{\pi}}(T)) \mid \widetilde{\G}^{\bullet}_0]=\varepsilon$, $P$-a.s., follows by the assumption that $\varepsilon \in [\varepsilon_{\min},\varepsilon_{\max}]$, defined in \eqref{eq epsilon min} and \eqref{eq epsilon max}. 
This implies that the risk constraint must be binding. 
Altogether, it holds that the optimal investment process $X^{\hat{\pi}}$ is unique and has to satisfy \eqref{eq lambda y}. \\
Next, let us show that the optimal trading strategy $\hat{\pi}$ is unique: 
Let us assume there exist $\pi^1,\pi^2 \in \Pi(\widetilde{\GG}^{\bullet},x)$ satisfying $X^{\pi^i}(t) = \widetilde{\EE}^\bullet[\hat{R} \mid \widetilde{\G}^{\bullet}_{t}]$, $i=1,2$. 
We define processes $N^i=(N^i_t)_{t \in [0,T]}$, $i=1,2$, by:  
$$ N^i_t \ = \ \int_{0}^{t} \pi^i_{u} \sigma_{u} \, d\widetilde{B}^{\bullet}_{u}, \qquad 0 \leq t \leq T. $$
$N^1$, $N^2$ are $\widetilde{P}^\bullet$-martingales w.\,r.\,t.\ the filtration $\widetilde{\GG}^{\bullet}$, so the difference $N^1-N^2$ is also a $\widetilde{P}^\bullet$-martingale. 
By $X^{\pi^1}(T)=\hat{R}=X^{\pi^2}(T)$, it follows that $N^1_T - N^2_T = 0$, and by its martingale property: $N^1-N^2 \equiv 0$. 
So, for its quadratic variation, it holds:
\begin{eqnarray*}
\langle N^1 - N^2 \rangle_T & = & \left\langle \int_{0}^{\cdot} \pi^1_{u} \sigma_{u} \, d\widetilde{B}^{\bullet}_{u} - \int_{0}^{\cdot} \pi^2_{u} \sigma_{u} \, d\widetilde{B}^{\bullet}_{u} \right\rangle_T \ = \ \int_{0}^{T} \left| \left(\pi^1_{u}-\pi^2_u\right) \sigma_{u}\right|^2 \, du \ = \ 0.
\end{eqnarray*}
Therefore, it must hold: $\pi^1 \equiv \pi^2 \equiv \hat{\pi}$.
\hfill{$\Box$} 

\begin{remark} \label{remark def z}
The optimization problem even has a solution, if we only assume (NA1) in the filtration $\widetilde{\GG}^{\bullet}$ since the martingale representation theorem for $\widetilde{\GG}^W$ holds by Assumption \ref{ass 3} (b) and for $\widetilde{\GG}^S$ it still holds under the weaker no arbitrage assumption (cf.\ \cite[Proposition 5.8.]{Fontana2013}). 
So, the process $\widetilde{Z}^{\bullet}$ may fail to be a true martingale and we cannot define $\widetilde{P}^\bullet$ as a probability measure and the process $\widetilde{B}^{\bullet}$ is only a drifted $P$-Brownian motion. 
Nevertheless, we obtain a similar result as in Theorem \ref{theorem optimal trading strategy g w tau}: 
The unique optimal trading strategy $\hat{\pi} \in \Pi(\widetilde{\GG}^{\bullet},x)$ and the corresponding wealth process $X^{\hat{\pi}}$ satisfy
$$ X^{\hat{\pi}}(t) \ = \ x + \int_{0}^{t} \hat{\pi}_{u} \sigma_{u} \, d\widetilde{B}^{\bullet}_{u} \ = \ (\widetilde{Z}^{\bullet})^{-1}_t \cdot \EE\left[ \widetilde{Z}^{\bullet}_T \hat{R} \, \left| \, \widetilde{\G}^{\bullet}_{t} \right.\right]. $$
\end{remark}

Let us end this section by giving an example for special utility and loss functions. 

\begin{example} \label{example1}
Let $U(k)=\ln k$ and $L(k)= - \frac3k$ be given. 
Then all properties of Definitions \ref{def utility function} and \ref{def loss function} are satisfied. 
Then we have 
$$ \tilde{U}_\lambda(k) \ = \ \ln k - \frac{3\lambda}k \qquad \text{and} \qquad \tilde{I}_\lambda(k) \ = \ \frac{1+\sqrt{1 + 12 \lambda k}}{2k}. $$ 
Now, the optimal terminal value for Problem \ref{problem1} is given by
$$ \hat{R} \ = \  \frac{1+\sqrt{1 + 12 \lambda^* \hat{y} \widetilde{Z}^\bullet_T}}{2\hat{y} \widetilde{Z}^\bullet_T}, $$
where $\hat{y}$ and $\lambda^*$ are such that equations \eqref{eq lambda y} are satisfied. \\
For the optimal wealth process $X^{\hat{\pi}}$ it holds that
$$ X^{\hat{\pi}}(t) \ = \ x + \int_0^t \hat{\pi}_u \, d\tilde{S}_u \ = \ x + \int_0^t \hat{\pi}_u \sigma_u \, d\widetilde{B}^\bullet_u \ = \ \EE \left[\left.\frac{1+\sqrt{1 + 12 \lambda^* \hat{y} \widetilde{Z}^\bullet_T}}{2\hat{y} \widetilde{Z}^\bullet_T} \, \right| \, \widetilde{\G}^{\bullet}_t \right]. $$
The output for the investor is given by
$$ u(\widetilde{\GG}^{\bullet},x) \ = \ \EE[U(\hat{R}) \mid \widetilde{\G}^{\bullet}_0] \ = \ \EE \left[\left. \ln \left(\frac{1+\sqrt{1 + 12 \lambda^* \hat{y} \widetilde{Z}^\bullet_T}}{2\hat{y} \widetilde{Z}^\bullet_T}\right) \, \right| \, \widetilde{\G}^{\bullet}_0 \right]. $$
%
%
Moreover, let us now assume that $\mu^i$ and $\sigma^i$, $i=1,2$, only depend on time and $\tilde{\theta} \equiv 0$ (i.\,e.\ $\widetilde{W}\equiv W$). 
Then the optimal wealth process is given by
$$ X^{\hat{\pi}}(t) \ = \ \frac1{2\hat{y}\widetilde{Z}^\bullet_t} \left(1 + \frac1{\sqrt{2\pi}} \int_{-\infty}^{\infty}(e^{-x^2}+12\lambda^*\hat{y}\widetilde{Z}^\bullet_te^{-(x-b/2)^2 + a+b^2/4})^{\frac12}  \, dx\right), $$
where $a:=- \frac{1}{2} \int_t^T {\left(||\theta_s||^2 \right)} ds$, $b:=-\sqrt{\int_t^T {\left(||\theta_s||^2 \right)} ds}$.
Moreover, $\hat{y}$ and $\lambda^*$ are such that the equations in \eqref{eq lambda y} are satisfied. 
The corresponding trading strategy is given by
\begin{eqnarray} \label{eq ex optimal trading}
\hat{\pi}_t & = & \frac{\mu^\bullet(t)}{\sigma^2(t)} \cdot \left(X^{\hat{\pi}}(t) + \frac{6 \lambda^*}{\hat{y} \sqrt{2\pi}} e^{a+b^2/4} \int_{-\infty}^\infty (e^{-x^2}+12\lambda^*\hat{y}\widetilde{Z}^\bullet_t e^{-(x-b/2)^2 + a+b^2/4})e^{-(x-b/2)^2} \, dx \right).
\end{eqnarray}
\end{example}

{\em Proof.}
Since the change point $\tau$ is known already at time zero, the drift and the volatility process are deterministic functions of time.
So we have:
\begin{eqnarray*} 
\tilde{\mu}^{\bullet}(t) & = & {\bf 1}_{\{t \leq \tau\}}\mu^{1}(t) + {\bf 1}_{\{t > \tau\}} \mu^{2}(t), \quad \sigma(t) \ := \ {\bf 1}_{\{t \leq \tau\}} \sigma^{1}(t) + {\bf 1}_{\{t > \tau\}}  \sigma^{2}(t).
\end{eqnarray*} 
Define the function $\Lambda^\bullet:[0,T] \to \RR$ by $\Lambda^\bullet(t)= \frac{\mu^\bullet(t)}{\sigma(t)}$, $t \in [0,T]$. 
The density $\widetilde{Z}^\bullet$ of the equivalent martingale measure as given in Assumption \eqref{ass 3} (c), can be expressed by
\begin{eqnarray*}
\widetilde{Z}^\bullet_T & = & \exp \left\{ -\int_0^T {\Lambda^\bullet_s} \, dW_s - \frac{1}{2} \int_0^T {(\Lambda^\bullet_s)^2} \, ds \right\} \ = \ \widetilde{Z}^\bullet_t \cdot \exp \left\{ -\int_t^T {\Lambda^\bullet_s} \, dW_s - \frac{1}{2} \int_t^T {(\Lambda^\bullet_s)^2} \, ds \right\} \\
& = & \widetilde{Z}^\bullet_t \cdot \exp(a + b \eta),
\end{eqnarray*}
where $a:=- \frac{1}{2} \int_t^T {(\Lambda^\bullet)^2(t)} ds$, $b:=-\sqrt{\int_t^T {(\Lambda^\bullet)^2(t)} ds}$ and $\eta$ is a standard Gaussian random variable independent of $\widetilde{\mathcal{G}}^\bullet_t$. 
The process $\widetilde{Z}^\bullet X^{\hat{\pi}}$ is a martingale with respect to $P$, so we have
\begin{eqnarray*}
\widetilde{Z}^\bullet_t X^{\hat{\pi}}_t & = & \EE[ \widetilde{Z}^\bullet_T X^{\hat{\pi}}_T \, | \, \widetilde{\mathcal{G}}^\bullet_t] \\
\Leftrightarrow \ X^{\hat{\pi}}_t & = & \EE \left[ \left. \frac{\widetilde{Z}^\bullet_T}{\widetilde{Z}^\bullet_t} \tilde{I}_{\lambda^*}(\hat{y} \widetilde{Z}^\bullet_T) \, \right| \, \widetilde{\mathcal{G}}^\bullet_t \right] 
\ = \ \EE \left[ \left. \frac{\widetilde{Z}^\bullet_T}{\widetilde{Z}^\bullet_t} \frac{1+\sqrt{1 + 12 \lambda^* \hat{y} \widetilde{Z}^\bullet_T}}{2\hat{y} \widetilde{Z}^\bullet_T} \, \right| \, \widetilde{\mathcal{G}}^\bullet_t \right].
\end{eqnarray*}
Following \cite{Gabih2005} we can use the representation
$$ \frac{c}{\widetilde{Z}^\bullet_t} \EE[g(\widetilde{Z}^\bullet_t,\eta) \, | \, \widetilde{\mathcal{G}}^\bullet_t] \ = \ \frac{c}{\widetilde{Z}^\bullet_t} \psi(\widetilde{Z}^\bullet_t) $$
with $\psi(z) = \EE[g(z,\eta)]$ for $z \in (0,+\infty)$, where $g$ is a measurable function and $c \in \RR$ is a constant, and derive the process $X^\pi$ in the way that
\begin{eqnarray*}
X^{\hat{\pi}}_t & = & \frac1{2\hat{y}\widetilde{Z}^\bullet_t} \cdot \left(1 + \EE\left[\left. \sqrt{1 + 12 \lambda^* \hat{y} \widetilde{Z}^\bullet_T} \, \right| \, \widetilde{\mathcal{G}}^\bullet_t \right] \right).
\end{eqnarray*}
Choose $g(z,x) = (1+12\lambda^*\hat{y}ze^{a+bx})^{\frac12}$ and with it we compute
\begin{eqnarray*}
\psi(z) & = & \EE[g(z,\eta)] \ = \ \frac1{\sqrt{2\pi}} \int_{-\infty}^{\infty}(1+12\lambda^*\hat{y}ze^{a+bx})^{\frac12} e^{-\frac12x^2}  \, dx \\
& = & \frac1{\sqrt{2\pi}} \int_{-\infty}^{\infty}(e^{-x^2}+12\lambda^*\hat{y}ze^{-(x-b/2)^2 + a+b^2/4})^{\frac12}  \, dx.
\end{eqnarray*}
Now, set $X^{\hat{\pi}}(t) = \frac1{2\hat{y}\widetilde{Z}^\bullet_t} (1+\psi(\widetilde{Z}^\bullet_t)) = F(\widetilde{Z}^\bullet_t,t)$ with
$$F(z,t) \ := \ \frac1{2\hat{y}z} \left(1 + \frac1{\sqrt{2\pi}} \int_{-\infty}^{\infty}(e^{-x^2}+12\lambda^*\hat{y}ze^{-(x-b/2)^2 + a+b^2/4})^{\frac12}  \, dx\right) , $$
it holds by It\^{o}'s formula that
\begin{eqnarray} \label{eq dynamics wealth process with f}
dX^{\hat{\pi}}(t) & = & F_t(\widetilde{Z}^\bullet_t,t) \, dt + F_z(\widetilde{Z}^\bullet_t,t) \, d\widetilde{Z}^\bullet_t + \frac12 F_{zz}(\widetilde{Z}^\bullet_t,t) \, d\widetilde{Z}^\bullet_td\widetilde{Z}^\bullet_t \nonumber \\
& = & \left(F_t(\widetilde{Z}^\bullet_t,t) + \frac12 F_{zz}(\widetilde{Z}^\bullet_t,t) (\widetilde{Z}^\bullet_t)^2 \Lambda^2(t)\right) \, dt - F_z(\widetilde{Z}^\bullet_t,t) \widetilde{Z}^\bullet_t \Lambda(t) \, dW_t,
\end{eqnarray}
where $F_z$, $F_{zz}$ and $F_t$ denote the partial derivatives of $F(z,t)$ with respect to $z$ and $t$. 
Comparing the coefficients in front of $dW_t$ in \eqref{eq wealth process v in g tau} and \eqref{eq dynamics wealth process with f}, we have that
\begin{equation*}
\hat{\pi}_t \sigma(t) \ = \ - F_z(\widetilde{Z}^\bullet_t,t) \widetilde{Z}^\bullet_t \Lambda^\bullet(t) \quad \iff \quad \hat{\pi}_t \ = \ -(\sigma(t))^{-1} \Lambda^\bullet(t) \widetilde{Z}^\bullet_t F_z(\widetilde{Z}^\bullet_t,t).
\end{equation*}
Let us compute the first order derivative of $F(z,t)$ w.\,r.\,t.\ $z$:
\begin{eqnarray*}
F_z(z,t) & = & -\frac1z \left(F(z,t) + \frac{6 \lambda^*}{\hat{y} \sqrt{2\pi}} e^{a+b^2/4} \int_{-\infty}^\infty (e^{-x^2}+12\lambda^*\hat{y}ze^{-(x-b/2)^2 + a+b^2/4})e^{-(x-b/2)^2} \, dx \right). 
\end{eqnarray*}
With this we get 
the expression in \eqref{eq ex optimal trading}.
\hfill{$\Box$}

\subsection{The progressively enlarged filtrations} \label{subsec g w}

We introduce two filtrations such that the change point $\tau$ is a stopping time w.\,r.\,t.\ these filtrations: 
\begin{itemize}
\item
$\mathbb{G}^W:=(\mathcal{G}^W_t)_{t \in [0,T]}$, where we $\mathcal{G}^W_t:=\bigcap_{s>t} (\mathcal{F}^{W}_s \vee \sigma(s \wedge \tau))$ represents the knowledge of the Brownian motion $W$ up to time $t$ plus the knowledge of the random time $\tau$ if it has occurred before time $t$,
\item
$\mathbb{G}^S:=(\mathcal{G}^S_t)_{t \in [0,T]}$, where $\mathcal{G}^S_t:=\bigcap_{s>t} (\mathcal{F}^S_s \vee \sigma(s \wedge \tau))$ represents the knowledge of the price process $S$ up to time $t$ plus the knowledge of the random time $\tau$ if it has occurred before time $t$. 
\end{itemize}
These are the smallest filtrations containing the filtration $\mathbb{F}^{W}$ and $\FF^S$, such that $\tau$ is a $\mathbb{G}^W$- and a $\GG^S$-stopping time, respectively. 
They are motivated by financial markets where an investor has insider information, e.\,g.\ he knows that the market parameters will change if the stock price will reach some benchmark. \\

Since it holds that $\GG^{S} \subseteq \GG^{W}$, the results for these two filtrations are very similar. 
In this setting, both initial $\sigma$-algebras are trivial. \\
We start with $\GG^{W}$ and make sure that the $\FF^{W}$-Brownian motions are still $\GG^W$-semimartingales. 

\begin{assumption} 
$ $  \label{ass 2} 
\begin{enumerate}
\item[(a)] There exists a $\mathbb{G}^W$-predictable process $\theta=(\theta_t)_{t \in [0,T]}$ and a $\mathbb{G}^W$-Brownian motion $W^*=(W^*_t)_{t \in [0,T]}$ such that:
$$ W_t \ = \ W^*_t + \int_0^t \theta_u \, du, \qquad \mbox{for all } t \in [0,T]. $$ 
Moreover, we define the $\mathbb{G}^W$-predictable process $\mu^W:=(\mu^W_{t})_{t \in [0,T]}$ by
\begin{eqnarray} 
\mu^W_{t} & := & {\bf 1}_{\{t \leq \tau\}} \left(\mu^{1}(t,\tilde{S}_{t}) + \sigma^{1}(t,\tilde{S}_{t}) \theta_{t} \right) + {\bf 1}_{\{t > \tau\}} \left(\mu^{2}(t,\tilde{S}_{t}) + \sigma^{2}(t,\tilde{S}_{t}) \theta_{t} \right), \label{eq mu for g} 
\end{eqnarray}
\item[(b)] 
The martingale representation property holds: 
Let $A^{W}$ be the $\mathbb{G}^W$-predictable compensator of the process $({\bf 1}_{\{\tau \leq t\}})_{t \in [0,T]}$ and let $N^{W}_\cdot:={\bf 1}_{\{\tau \leq \cdot\}} - A^{W}_\cdot$ be the corresponding compensated martingale w.\,r.\,t.\ the filtration $\mathbb{G}^W$. 
Then every $\mathbb{G}^W$-local martingale $L^W=(L^W_{t})_{t \in [0,T]}$ admits the representation
\begin{equation} \label{eq mrt g}
L^W_t \ := \ L^W_{0} + \int_{0}^{t} \varphi^W_{u} \, dW^*_{u} + \int_0^t \psi^W_u \, dN^{W}_u, \qquad 0 \leq t \leq T, 
\end{equation}
for some $\mathbb{G}^W$-predictable processes $\varphi^W=(\varphi^W_{t})_{t \in [0,T]}$ satisfying $\int_{0}^{T} (\varphi^W_{t})^{2} dt < +\infty$, $P$-a.s., and $\psi^W=(\psi^W_{t})_{t \in [0,T]}$ satisfying $\int_{0}^{T} |\psi^W_{t}| |dA^{W}_t| < +\infty$, $P$-a.s.
\item[(c)] 
It holds (NFLVR) in the filtration $\mathbb{G}^W$ and the process $Z^W=(Z^W_t)_{t \in [0,T]}$, defined by $Z^W_t:=\E(-\int\mu^W/\sigma dW^* + \int \upsilon^W dN^W)_t$ for some $\GG^W$-predictable process $\upsilon^W=(\upsilon^W_t)_{t \in [0,T]}$ satisfying\footnote{Define $\Delta N^W_t:=N^W_t-N^W_{t-}$ for all $t \in [0,T]$.} $\upsilon^W \Delta N^W > -1$ and $\int_0^T |\upsilon^W_t||dA^W_t|<+\infty$, $P$-a.s., is a true $P$-martingale w.\,r.\,t.\ $\mathbb{G}^W$. 
Moreover, define $P^W \sim P$ as a probability measure such that $\frac{dP^W}{dP} = Z^W_T$. 
\end{enumerate}
\end{assumption}

Let us give a short explanation to these assumptions. 

\begin{remark}  
$ $\label{remark ass in g w}
\begin{enumerate}
\item[(a)] 
Part (a) of this assumption is trivially satisfied, if the $\mathbb{F}^{W}$-Brownian motion $W$ and the random time $\tau$ are independent. 
In this case we have $\theta \equiv 0$. 
It is also (but not trivially) satisfied, if $\tau$ satisfies the density hypothesis, cf.\ \cite[Proposition 5.9]{ElKaroui2010}, or if $\tau$ is obtained through the canonical construction, cf.\ \cite{Bielecki2002}.  
In this case, the $\mathbb{F}^{W}$-Brownian motion $W$ remains a Brownian motion w.\,r.\,t.\ the filtration $\mathbb{G}^W$, hence $\theta \equiv 0$. 
Moreover, it is also satisfied if $\tau$ is an honest time\footnote{A random time $\tau$ is called {\em honest} if for every $t$ there exists an $\F_t$-measurable random variable $\sigma_t$ such that $\tau=\sigma_t$ on $\{\tau < t\}$, cf.\ \cite[Definition 8.10.]{nikeghbali}.}, cf.\ \cite[Chapter V]{Jeulin1980}.
\item[(b)] Conditions for the second assumption to hold can be found in \cite{Jeanblanc2015}.  
\item[(c)] 
Part (c) of the assumption is satisfied if e.\,g.\ it holds that $\EE[Z^W_T]=1$. 
Moreover, we have that $\int_{0}^{T}(\mu^W_{u}/\sigma_{u})^{2} du < +\infty$, $P$-a.s., cf.\ \cite[Proposition 3.3.]{Fontana2013}.
\end{enumerate}
\end{remark}

Now, let us repeat (cf.\ \cite[Proposition 3.1.]{Fontana2013}) that under Assumptions \ref{ass 1} and \ref{ass 2} (a) there exists a unique continuous $\mathbb{G}^W$-semimartingale $\tilde{S}=(\tilde{S}_{t})_{t \in [0,T]}$ which is a solution to the SDE \eqref{eq tilde s general} on $(\Omega,\mathcal{A},\mathbb{G}^W,P)$. 
$\tilde{S}$ has the following representation w.\,r.\,t.\ the filtration $\mathbb{G}^W$:
\begin{equation} \label{eq repr x for g}
\tilde{S}_{t} \ = \ \int_{0}^{t} \tilde{\mu}_{u} \, du + \int_{0}^{t} \sigma_{u} \, dW^*_{u}, \qquad \mbox{for all } t \in [0,T].
\end{equation}


Let us now consider the representation of the price process and of local martingales w.\,r.\,t.\ the filtration $\GG^S$.

\begin{lemma} {\em (cf.\ \cite[Lemma 3.6., Proposition 3.10.]{Fontana2013})} \label{lemma representation s on g s}
Suppose Assumptions \ref{ass 1} and \ref{ass 2} hold true.\footnote{Note that for statement (ii) it is enough to claim (NA1) instead of (NFLVR) in the filtration $\GG^S$.}
\begin{enumerate}
\item[(i)]
The process $\tilde{S}=(\tilde{S}_{t})_{t \in [0,T]}$ has the following representation w.r.t.\ the filtration $\mathbb{G}^{S}$:
\begin{equation} \label{eq repr x for g s}
\tilde{S}_{t} \ = \ \int_{0}^{t} \mu^S_{u} \, du + \int_{0}^{t} \sigma_{u} \, dW^S_{u}, \qquad \mbox{for all } t \in [0,T],
\end{equation}
where the $\mathbb{G}^{S}$-predictable process $\mu^S:=(\mu^S_{t})_{t \in [0,T]}$ is defined as
\begin{eqnarray} \label{eq mu for g s}
\mu^S_{t} & := & {\bf 1}_{\{t \leq \tau\}} \left(\mu^{1}(t,\tilde{S}_{t}) + \sigma^{1}(t,\tilde{S}_{t}) \, ^{p}\theta_{t} \right) + {\bf 1}_{\{t > \tau\}} \left(\mu^{2}(t,\tilde{S}_{t}) + \sigma^{2}(t,\tilde{S}_{t}) \,  ^{p}\theta_{t} \right),
\end{eqnarray}
where $^{p}\theta$ denotes the $\mathbb{G}^{S}$-predictable projection of $\theta$, the process $W^S=(W^S_{t})_{t \in [0,T]}$ is a $\mathbb{G}^{S}$-Brownian motion and $\sigma=(\sigma_{t})_{t \in [0,T]}$ is given in \eqref{eq sigma general}. 
We have that $\mu^S=\,^p\mu^W$. 
\item[(ii)]
Let $A^{S}$ be the $\mathbb{G}^S$-predictable compensator of the process $({\bf 1}_{\{\tau \leq t\}})_{t \in [0,T]}$ and let $N^{S}$ be the corresponding compensated martingale w.\,r.\,t.\ the filtration $\mathbb{G}^S$. 
Every $\GG^S$-local martingale $L^S=(L^S_t)_{t \in [0,T]}$ has the form
$$ L^S_t \ := \ L^S_{0} + \int_{0}^{t} \varphi^S_{u} \, dW^S_{u} + \int_{0}^{t}\psi^S_{u} \, dN^{S}_{u}, \qquad 0 \leq t \leq T, $$
for some $\mathbb{G}^{S}$-predictable processes $\varphi^S=(\varphi^S_{t})_{t \in [0,T]}$ satisfying $\int_{0}^{T} (\varphi^S_{t})^{2} dt < +\infty$, $P$-a.s., and $\psi^S=(\psi^S_{t})_{t \in [0,T]}$ satisfying $\int_{0}^{T} |\psi^S_{t}| |dA^{S}_{t}| < +\infty$, $P$-a.s.
\end{enumerate}
\end{lemma}

By Assumption \ref{ass 2} (c), (NFLVR) also holds in the smaller filtration $\GG^{S}$, so the $\GG^{S}$-adapted process $Z^{S}=(Z^{S}_{t})_{t \in [0,T]}$ defined by $Z^{S}_t=\E(\int \frac{\mu^{S}}{\sigma} dW^{S}+\int \upsilon^S dN^S)_t$ for some $\GG^S$-predictable process $\upsilon^S=(\upsilon^S_t)_{t \in [0,T]}$ satisfying $\upsilon^S \Delta N^S>-1$ and $\int_0^T|\upsilon^S_t||dA^S_t|<+\infty$, $P$-a.s., is a true martingale and we can define $P^S \sim P$ as a probability measure such that $ \frac{dP^S}{dP} = Z^{S}_T$. \\
From now on if we state results holding for both filtrations, we simply write $\bullet \in \{W,S\}$, so e.\,g.\ $\GG^{\bullet}$ for $\GG^{W}$ and $\GG^{S}$ or $\G^{\bullet}_t$ for $\G^{W}_t$ and $\G^{S}_t$, $t \in [0,T]$, respectively. 
Note, that $W^W:=W^*$. \\
Moreover, by $\EE^\bullet$ we denote the expectation w.\,r.\,t.\ $P^\bullet$. \\

Let us now define portfolio processes for this subsection. 
In the present setting it will be no longer self-financing and we will give an interpretation for this directly after the definition.

\begin{definition} \label{def trading strategy g w}
Consider $\mathbb{G}^\bullet$-predictable processes $\pi=(\pi_t)_{t \in [0,T]}$ satisfying $\int_{0}^{T} (\pi_{u}\sigma_{u})^{2} du < +\infty$, $P$-a.s., and $\psi=(\psi_t)_{t \in [0,T]}$ satisfying $\int_0^T {|\psi_t| \, |dA^{\bullet}_t|} < +\infty$, $P$-a.s.
We call the couple $\zeta=(\pi,\psi)$ {\em trading strategy} and define the corresponding wealth process by
\begin{equation} \label{eq wealth process v in g w}
X^{\zeta}(t) \ = \ x + \int_{0}^{t} \pi_{u} \, d\tilde{S}_{u} + \int_0^t \psi_u \, dN^{\bullet}_u \ = \ x + \int_{0}^{t} \pi_{u}\mu^\bullet_{u}\, du + \int_{0}^{t} \pi_{u} \sigma_{u} \, dW^\bullet_{u} + \int_0^t \psi_u \, dN^{\bullet}_u, \ 0 \leq t \leq T,
\end{equation}
where the initial capital $x \in \RR$ is assumed to be greater than zero. 
$\zeta$ is called {\em admissible} if the corresponding process $X^\zeta(t)$ is non-negative for all $t \in [0,T]$, $P$-a.s. 
The set of all admissible trading strategies with initial capital $x$ is denoted by $\Pi(\GG^\bullet,x)$, the set of corresponding wealth processes by $\X(\GG^\bullet,x)$.
\end{definition}

\begin{remark}
\begin{enumerate}
\item[]
\item[(a)]
By the assumption of (NFLVR), equation \eqref{eq wealth process v in g w} is well-defined, since by the Cauchy-Schwarz inequality it holds that $\int_0^T \pi_t \mu^\bullet_t \, dt < +\infty$, $P$-a.s. 
\item[(b)]
The process $\eta_t=\int_0^t \psi_u \, dN^{\bullet}_u$, $t \in [0,t]$, can be interpreted as a nontradable asset (cf.\ \cite{Mania2010}) or as additional endowment of the agent. 
So in this case, we consider strategies which are not necessarily self-financing and every contingent claim which should be hedged will carry an intrinsic risk. 
The condition $\int_0^T {|\psi_t| \, |dA^{\bullet}_t|} < +\infty$, $P$-a.s., ensures that the trading strategy is {\em mean-self-financing} in the sense that
$$ \EE[\eta_T \mid \G^\bullet_t] \ = \ \eta_t, \qquad t \in [0,T], $$
i.\,e.\ that $\eta$ is a martingale w.\,r.\,t.\ $\GG^\bullet$, also cf.\ \cite{Foellmer1991}. \\
The market becomes complete if we define a two-dimensional stock price process $S^*$ by $S^*=(S,N^{\bullet})$, i.\,e.\ we add the process $N^{\bullet}$ to the market, cf.\ \cite{Foellmer1991}.
\end{enumerate}
\end{remark}

Define the process $B^\bullet=(B^\bullet_{t})_{t \in [0,T]}$ by 
$$ B^\bullet_{t} \ = \ B^\bullet_{t} + \int_{0}^{t} \frac{\mu^\bullet_{u}}{\sigma_{u}} \, du, \qquad 0 \leq t \leq T, $$
which is a Brownian motion under $P^\bullet$ w.\,r.\,t.\ the filtration $\mathbb{G}^\bullet$.
So, the process $\tilde{S}$ is of the form
\begin{eqnarray*}
\tilde{S}_{t} & = & \int_{0}^{t} \mu^\bullet_{u} \, du + \int_{0}^{t} \sigma_{u} \, dW^\bullet_{u}\ = \ \int_{0}^{t} \sigma_{u} \, dB^\bullet_{u}, \qquad \mbox{for all } t \in [0,T],
\end{eqnarray*}
and the wealth process $X^{\zeta}$ admits the representation
\begin{eqnarray*}
X^{\zeta}(t) & = & x+ \int_{0}^{t} \pi_{u} \sigma_{u} \, dB^\bullet_{u} + \int_0^t \psi_u \, dN^{\bullet}_u, \qquad \mbox{for all } t \in [0,T].
\end{eqnarray*}
Since $\langle Z^\bullet, N^{\bullet}\rangle =0$, $Z^\bullet N^{\bullet}$ is a $P$-martingale and therefore $N^{\bullet}$ is a $P^\bullet$-martingale. \\

To obtain unique solutions for the portfolios, let us first define risk minimizing strategies, which were introduced e.g.\ by \cite{Foellmer1986}.

\begin{definition}[Risk-minimizing strategy] \label{def risk minimizing strategy}
A trading strategy $\zeta^*=(\pi^*,\psi^*)$ is called {\em risk-minimizing} if for any other trading strategy $\zeta=(\pi,\psi)$ it holds that
$$ \EE\left[\left(\int_0^t \psi^*_u \, dN^{\bullet}_u\right)^2\right] \ \leq \ \EE\left[\left(\int_0^t \psi_u \, dN^{\bullet}_u\right)^2\right], $$
i.e.\ $\zeta^*$ minimizes the variance of the random variable $\eta_T:=\int_0^T \psi_u \, dN^{\bullet}_u$. 
\end{definition}

We give the main result of this subsection:

\begin{theorem} \label{thm g w}
Suppose Assumptions \ref{ass 1} and \ref{ass 2} hold true. 
Moreover, let $\EE[Z^\bullet_T \cdot \tilde{I}_\lambda(yZ^\bullet_T)]<+\infty$ for all $y \in (0,+\infty)$. 
Then the optimal terminal wealth $\hat{R}$ for Problem \ref{problem1} is given by
$$ \hat{R} \ = \ \tilde{I}_{\lambda^*}\left( \hat{y}Z^\bullet_T \right), $$
where $\hat{y} \in (0,+\infty)$ and $\lambda^* \geq 0$ are such that
\begin{equation} \label{eq y lambda2}
\left\{ \begin{array} {rcl}  \EE\left[Z^\bullet_T \cdot \tilde{I}_{\lambda^*}(\hat{y}Z^\bullet_T)\right] & =& x, \\ 
\EE\left[L(-\hat{R})\right] & = & \varepsilon. \end{array} \right. 
\end{equation}
The unique optimal risk-minimizing trading strategy $\hat{\zeta}=(\hat{\pi},\hat{\psi}) \in \Pi(\mathbb{G}^{\bullet},x)$ and the corresponding wealth process $X^{\hat{\zeta}}$ satisfy
$$ X^{\hat{\zeta}}(t) \ = \ x+ \int_{0}^{t} \hat{\pi}_{u} \sigma_{u} \, dB^\bullet_{u} + \int_0^t \hat{\psi}_u \, dN^{\bullet}_u \ = \ \EE^\bullet\left[\left. \hat{R} \, \right| \, \mathcal{G}^\bullet_{t}\right]. $$
\end{theorem}

To prove the theorem, we first state the martingale representation theorem under $P^\bullet$.

\begin{lemma} \label{lemma mrt g w}
Under the assumptions of Theorem \ref{thm g w}, every local martingale $M^\bullet=(M^\bullet_{t})_{t \in [0,T]}$ under $P^\bullet$ w.\,r.\,t.\ the filtration $\mathbb{G}^\bullet$ is of the form
\begin{equation} \label{eq mrt for hat p in g 2} 
M^\bullet_{t} \ = \ M^\bullet_0 + \int_0^t \pi_u \sigma_u \, dB^\bullet_u + \int_0^t \bar{\psi}^\bullet_u \, dN^{\bullet}_u, \qquad 0 \leq t \leq T,
\end{equation}
for some $\mathbb{G}^\bullet$-predictable stochastic processes $\pi=(\pi_t)_{t \in [0,T]}$ and $\bar{\psi}^\bullet=(\bar{\psi}^\bullet_t)_{t \in [0,T]}$ satisfying $\int_{0}^{T} (\pi_{u}\sigma_{u})^{2} du < +\infty$, $P$-a.s., and $\int_0^T {|\bar{\psi}^\bullet_t| |dA^{\bullet}_t|} dt < +\infty$, $P$-a.s.
\end{lemma}

{\em Proof.}
First, let us show that the process $Z^\bullet M^\bullet=(Z^\bullet_tM^\bullet_t)_{t \in [0,T]}$ is a $P$-local martingale. 
Since $M^\bullet$ is a local $P^\bullet$-martingale, there exists an increasing sequence of stopping times $(T_n)_{n \in \NN}$ satisfying $T_n \to +\infty$. 
Now it holds for any $0 \leq s \leq t \leq T$ and $n \in \NN$ by Bayes' formula:
$$ \EE\left[Z^\bullet_{t \wedge T_n} M^\bullet_{t \wedge T_n} \, | \, \mathcal{G}^{\bullet}_{s}\right] \ = \ Z^\bullet_{s \wedge T_n} \EE^\bullet\left[ M^\bullet_{t \wedge T_n} \, | \, \mathcal{G}^{\bullet}_{s}\right] \ = \ Z^\bullet_{s \wedge T_n} M^\bullet_{s \wedge T_n}. $$
By the $\GG^\bullet$-martingale representation property (as given in \eqref{eq mrt g}), the process $L^\bullet:=Z^\bullet M^\bullet$ has the form
$$ L^\bullet_t \ = \ L^\bullet_{0} + \int_{0}^{t} \varphi^\bullet_{u} \, dW^\bullet_{u} + \int_{0}^{t}\psi^\bullet_{u} \, dN^{\bullet}_{u}, \qquad 0 \leq t \leq T, $$
for some $\mathbb{G}^{S}$-predictable processes $\varphi^\bullet=(\varphi^\bullet_{t})_{t \in [0,T]}$ satisfying $\int_{0}^{T} (\varphi^\bullet_{t})^{2} dt < +\infty$, $P$-a.s., and $\psi^\bullet=(\psi^\bullet_{t})_{t \in [0,T]}$ satisfying $\int_{0}^{T} |\psi^\bullet_{t}| |dA^{\bullet}_{t}| < +\infty$, $P$-a.s.
Then, we have $M^\bullet=L^\bullet (Z^\bullet)^{-1}$, where the dynamics of $(Z^\bullet)^{-1}$ by It\^{o}'s rule is
\begin{eqnarray*}
d(Z^\bullet)^{-1}_t & = & -(Z^\bullet)^{-1}_t \, d\left(-\int_{0}^{t} \frac{\mu^\bullet_{u}}{\sigma_{u}} \, dW^\bullet_{u} - \frac12 \int_{0}^{t} \left(\frac{\mu^\bullet_{u}}{\sigma_{u}} \right)^{2} \, du + \int_0^t \upsilon^\bullet_u \, dN^\bullet_u \right) + \frac12 (Z^\bullet)^{-1}_t \left(\frac{\mu^\bullet_{t}}{\sigma_{t}} \right)^{2} \, dt \\
& = & \frac{\mu^\bullet_{t}}{\sigma_{t}} (Z^\bullet)^{-1}_t \, dW^\bullet_t + \left(\frac{\mu^\bullet_{t}}{\sigma_{t}} \right)^{2}(Z^\bullet)^{-1}_t \, dt + \upsilon^\bullet_t (Z^\bullet)^{-1}_t \, dN^\bullet_t.
\end{eqnarray*}
Using It\^{o}'s product rule, the dynamics of $M^\bullet$ is given by:
\begin{eqnarray*}
dM^\bullet_t & = & d\left(L^\bullet (Z^\bullet)^{-1}\right)_t \\
& = & L^\bullet_t \, d(Z^\bullet)^{-1}_t + (Z^\bullet)^{-1}_t \, dL^\bullet_t + d \langle L^\bullet, (Z^\bullet)^{-1} \rangle_t \\
& = & Z^\bullet_{t}M^\bullet_{t} \cdot \left(\frac{\mu^\bullet_{t}}{\sigma_{t}} (Z^\bullet)^{-1}_t \, dW^\bullet_t + \left(\frac{\mu^\bullet_{t}}{\sigma_{t}}\right)^{2}(Z^\bullet)^{-1}_t \, dt + \upsilon^\bullet_t (Z^\bullet)^{-1}_t \, dN^\bullet_t \right) + (Z^\bullet)^{-1}_t \cdot \left( \varphi^\bullet_t \, dW^\bullet_t + \psi_t \, dN^{\bullet}_t \right) \\ 
& & + (Z^\bullet)^{-1}_t \varphi^\bullet_t \frac{\mu^\bullet_{t}}{\sigma_{t}} \, dt + (Z^\bullet)^{-1}_t \psi^\bullet_t \frac{\mu^\bullet_{t}}{\sigma_{t}} \, d\underbrace{\langle W^\bullet, N^{\bullet} \rangle_t}_{= \, 0} \\
& = & M^\bullet_t \frac{\mu^\bullet_{t}}{\sigma_{t}} \, dW^\bullet_t + M^\bullet_t \left(\frac{\mu^\bullet_{t}}{\sigma_{t}}\right)^2 \, dt + M^\bullet_t \upsilon^\bullet_t \, dN^\bullet_t + (Z^\bullet)^{-1}_t \varphi^\bullet_t \, dW^\bullet_t + (Z^\bullet)^{-1}_t \psi^\bullet_t \, dN^{\bullet}_t + (Z^\bullet)^{-1}_t \varphi^\bullet_t \frac{\mu^\bullet_{t}}{\sigma_{t}} \, dt \\
& = & \left(M^\bullet_{t} \frac{\mu^\bullet_{t}}{\sigma_{t}} + (Z^\bullet)^{-1}_t \varphi^\bullet_t\right) \cdot \left(dW^\bullet_t + \frac{\mu^\bullet_{t}}{\sigma_{t}} \, dt \right) + \left( M^\bullet_t \upsilon^\bullet_t + (Z^\bullet)^{-1}_t \psi^\bullet_t\right) \, dN^{\bullet}_t \\
& = & \left(M^\bullet_{t} \frac{\mu^\bullet_{t}}{\sigma_{t}} + (Z^\bullet)^{-1}_t \varphi^\bullet_t\right) \, dB^\bullet_t + \left( M^\bullet_t \upsilon^\bullet_t + (Z^\bullet)^{-1}_t \psi^\bullet_t\right) \, dN^{\bullet}_t .
\end{eqnarray*}
Set $\bar{\varphi}^\bullet_{t}:=M^\bullet_{t} \frac{\mu^\bullet_{t}}{\sigma_{t}} + (Z^\bullet)^{-1}_t \varphi^\bullet_t$, and $\bar{\psi}^\bullet_t:=M^\bullet_t \upsilon^\bullet_t + (Z^\bullet)^{-1}_t \psi^\bullet_t$, then $M^\bullet$ is of the form
\begin{equation*}  
M^\bullet_{t} \ = \ M^\bullet_{0} + \int_{0}^{t} \bar{\varphi}^\bullet_{u} \, dB^\bullet_{u} + \int_0^t \bar{\psi}^\bullet_u \, dN^{\bullet}_u, \qquad 0 \leq t \leq T, 
\end{equation*}
where $\bar{\varphi}^\bullet=(\bar{\varphi}^\bullet_{t})_{t \in [0,T]}$ and $\bar{\psi}^\bullet=(\bar{\psi}^\bullet_{t})_{t \in [0,T]}$ are $\mathbb{G}^\bullet$-predictable processes satisfying $\int_{0}^{T} (\bar{\varphi}^\bullet_{t})^{2} dt < +\infty$, $P$-a.s., and $\int_{0}^{T} |\bar{\psi}^\bullet_{t}| |dA^{\bullet}| < +\infty$, $P$-a.s., which follows from Remark \ref{remark ass in g w} (b) and the path continuity of $Z^\bullet$ and $M^\bullet$, cf.\ \cite{Karatzas1991}.
Set $\pi:=\bar{\varphi}^\bullet/\sigma$, then it holds that $\int_{0}^{T} (\pi_{u}\sigma_{u})^{2} du < +\infty$, $P$-a.s.
\hfill{$\Box$} \\

Let us now give the proof of the main result. \\

{\bf Proof of Theorem \ref{thm g w}.} 
The existence of $\hat{y} \in (0,+\infty)$ and $\lambda^* \geq 0$ such that equations \eqref{eq y lambda2} hold is proven in Appendix \ref{app lagrangian}. 
We show that there exists a trading strategy $\hat{\zeta}=(\hat{\pi},\hat{\psi}) \in \Pi(\mathbb{G}^{\bullet},x)$ such that for the corresponding wealth process $X^{\hat{\zeta}}$ it holds:
$$ X^{\hat{\zeta}}(t) \ = \ \EE^\bullet\left[ \hat{R} \, | \, \mathcal{G}^{\bullet}_{t}\right]. $$
Define $M(t) = \EE^\bullet\left[\hat{R} \, | \, \mathcal{G}^{\bullet}_{t}\right]$, $0 \leq t \leq T$. 
Because for $0 \leq s \leq t$ we have
\begin{eqnarray*}
\EE^\bullet\left[M(t) \, | \, \mathcal{G}^{\bullet}_{s}\right] & = & \EE^\bullet\left[\EE^\bullet\left[\hat{R} \, | \, \mathcal{G}^{\bullet}_{t}\right] \, | \, \mathcal{G}^{\bullet}_{s}\right] \ = \ \EE^\bullet\left[\hat{R} \, | \, \mathcal{G}^{\bullet}_{s}\right] \ = \ M(s),
\end{eqnarray*}
$M$ is a $P^\bullet$-martingale w.\,r.\,t.\ the filtration $\mathbb{G}^{\bullet}$. 
By Lemma \ref{lemma mrt g w}, we obtain:
$$ M(t) \ = \ M(0) + \int_{0}^{t} \hat{\pi}_{u} \sigma_{u} \, dB^\bullet_{u} + \int_0^t \hat{\psi}_u \, dN^{\bullet}_u, \qquad 0 \leq t \leq T, $$
for some $\mathbb{G}^\bullet$-predictable processes $\hat{\pi}=(\hat{\pi}_t)_{t \in [0,T]}$ and $\hat{\psi}=(\hat{\psi}_{t})_{t \in [0,T]}$ satisfying $\int_{0}^{T} (\hat{\pi}_{t}\sigma_t)^2 dt < +\infty$, $P$-a.s., and $\int_{0}^{T} |\hat{\psi}_{t}| |dA^{\bullet}_t| < +\infty$, $P$-a.s., and with $M(0)=\EE^\bullet[\hat{R}] = x$. 
So, it holds: $X^{\hat{\zeta}}(t) = M(t) = \EE^\bullet[\hat{R} \mid \mathcal{G}^{\bullet}_{t}] \geq 0$, since $\hat{R}$ is non-negative.
therefore, $\hat{\zeta}$ is indeed in $\Pi(\mathbb{G}^\bullet,x)$. \\
For the optimality, let us consider another $\zeta \in \Pi(\mathbb{G}^{\bullet},x)$. 
By the inequality \eqref{eq inequality u} it holds:
\begin{eqnarray*}
\tilde{U}_{\lambda^*}(\hat{R}) \ = \ \tilde{U}_{\lambda^*}(\tilde{I}_{\lambda^*}(\hat{y}Z^\bullet_T)) \ = \ \tilde{U}_{\lambda^*}(X^{\hat{\zeta}}(T)) & \geq & \tilde{U}_{\lambda^*}(X^\zeta(T)) + \hat{y}Z^\bullet_T \cdot \hat{R} - \hat{y}Z^\bullet_T \cdot X^\zeta(T).
\end{eqnarray*}
Taking expectations on both sides and since $\EE^\bullet[X^\zeta(T)]\leq \ x$, it follows:
\begin{eqnarray*}
\EE\left[\tilde{U}_{\lambda^*}(X^{\hat{\zeta}}(T))\right] & \geq & \EE \left[\tilde{U}_{\lambda^*}(X^\zeta(T)) + \hat{y}Z^\bullet_T \cdot \hat{R} - \hat{y}Z^\bullet_T \cdot X^\zeta(T) \right] \\
 & & = \EE \left[\tilde{U}_{\lambda^*}(X^\zeta(T))\right] + \hat{y} \left( \EE\left[Z^\bullet_T \cdot \hat{R}\right] - \EE\left[Z^\bullet_T \cdot X^\zeta(T) \right] \right) \\
 & & = \EE \left[\tilde{U}_{\lambda^*}(X^\zeta(T))\right] + \hat{y}\left(\EE^\bullet\left[\hat{R}\right] - \EE^\bullet\left[X^\zeta(T) \right]\right) \ \geq \ \EE \left[\tilde{U}_{\lambda^*}(X^\zeta(T))\right].
\end{eqnarray*}
Therefore, $\hat{\zeta}$ is the optimal solution for Problem \ref{problem1}. \\
For the uniqueness of $\hat{\zeta}$, let us show that the optimal investment process $X^{\hat{\zeta}}$ is unique: 
For that, let us suppose there are $\zeta^1,\zeta^2 \in \Pi(\GG^{\bullet},x)$ such that $X^{\zeta^1}, X^{\zeta^2}$ are optimal for Problem \ref{problem1}.
Then we define $\zeta^3$ such that $X^{\zeta^3}=\frac12(X^{\zeta^1} + X^{\zeta^2})$. 
By the convexity of the loss function $L$ and the function $U^-$ we have that $X^{\zeta^3} \in \X(\GG^{\bullet},x)$. 
But on the other hand it holds due to the concavity of $U$ that
\begin{eqnarray*}
\EE \left[U(X^{\zeta^3}(T))\right] & = & \EE \left[U(1/2(X^{\zeta^1}(T)+X^{\zeta^2}(T)))\right] \\ 
& \geq & \frac12 \EE \left[U(X^{\zeta^1}(T))+U(X^{\zeta^2}(T)))\right] \ = \ \EE \left[U(X^{\zeta^1}(T))\right],
\end{eqnarray*}
where we have equality in the second line due to the optimality of $X^{\zeta^1}$ and $X^{\zeta^2}$. 
Therefore, we must have that $X^{\zeta^1}(T)=X^{\zeta^2}(T)$, $P$-a.s. \\
Next, we show that the optimal investment process has to satisfy \eqref{eq y lambda2}. 
First, let us start with $\EE^\bullet[X^{\hat{\zeta}}(T)] = x$. 
Suppose that $\EE^\bullet[X^{\hat{\zeta}}(T)] = 0$. 
Obviously we have that $X^{\hat{\zeta}}(T)\equiv 0$, since $\hat{\zeta} \in \Pi(\GG^{\bullet},x)$. 
Let us consider the trading strategy $\tilde{\zeta}=(\tilde{\pi},\tilde{\psi}) \equiv (0,0)$. 
It holds that $X^{\tilde{\zeta}}(t)=x$, $t \in [0,T]$. 
Especially we have that $X^{\tilde{\zeta}}(T) > X^{\hat{\zeta}}(T)$, $P$-a.s., and by concavity of $U$ and convexity of $L$ and $U^-$ it holds that $X^{\tilde{\zeta}} \in \X(\GG^{\bullet},x)$ and 
$$ \EE \left[U(X^{\tilde{\zeta}}(T))\right] \ > \ \EE \left[U(X^{\hat{\zeta}}(T))\right], $$
which contradicts the optimality of $X^{\hat{\zeta}}$. \\
Now, suppose that $\EE^\bullet[X^{\hat{\pi}}(T)] = a < x$. 
Choose a trading strategy $\zeta^*$ such that $X^{\zeta^*}=\frac{x}{a}X^{\hat{\zeta}}$. 
It holds that $\EE^\bullet[X^{\zeta^*}(T)]=x$ and $X^{\zeta^*}>X^{\hat{\zeta}}$. 
By the monotonicity of $U$, $L$ and $U^-$ we have that $X^{\zeta^*} \in \X(\GG^{\bullet},x)$ and $\EE[U(X^{\zeta^*}(T))] > \EE[U(X^{\hat{\zeta}}(T))]$; a contradiction to the optimality of $X^{\hat{\zeta}}$. \\
Last, the condition $\EE[L(-X^{\hat{\zeta}}(T))]=\varepsilon$ follows by the assumption that $\varepsilon \in [\varepsilon_{\min},\varepsilon_{\max}]$, defined in \eqref{eq epsilon min} and \eqref{eq epsilon max} which implies that the risk constraint is binding. 
Altogether, it holds that the optimal investment process $X^{\hat{\zeta}}$ is unique and has to satisfy \eqref{eq y lambda2}. \\
Next, let us show that the optimal risk-minimizing trading strategy $\hat{\zeta}$ is unique: 
Let us assume there exist $\zeta^1=(\pi^1,\psi^1),\zeta^2=(\pi^2,\psi^2) \in \Pi(\mathbb{G}^{\bullet},x)$ satisfying $X^{\zeta^i}(t) = \EE^\bullet[\hat{R} \mid \mathcal{G}^{\bullet}_{t}]$, $i=1,2$. 
We first show that $\pi^1 \equiv \pi^2$. 
It holds:
$$ x + \int_0^T \psi^2_u \, dN^{\bullet}_u \ = \ \hat{R} - \int_0^T \pi^2_u \sigma_u \, dB^\bullet_u \ = \ x + \int_0^T (\pi^1_u - \pi^2_u) \sigma_u \, dB^\bullet_u + \int_0^T \psi^1_u \, dN^{\bullet}_u. $$
Since $N^{\bullet}$ and $B^\bullet$ are orthogonal, it follows that
\begin{eqnarray*}
\EE\left[\left(\int_0^T \psi^2_u \, dN^{\bullet}_u\right)^2\right] & = & \EE\left[\left(\int_0^T (\pi^1_u - \pi^2_u) \sigma_u \, dB^\bullet_u\right)^2\right] + \EE\left[\left(\int_0^T \psi^1_u \, dN^{\bullet}_u\right)^2\right] \\
& = & \EE\left[\int_0^T (\pi^1_u - \pi^2_u)^2 \sigma^2_u \, du\right] + \EE\left[\left(\int_0^T \psi^1_u \, dN^{\bullet}_u\right)^2\right], 
\end{eqnarray*}
which is minimized if and only if $\pi^1 \equiv \pi^2$. \\
Last, we obtain by the uniqueness of the optimizing value processes $X^{\zeta^1}$, $X^{\zeta^2}$ that $X^{\zeta^1}(t) = \EE^\bullet[\hat{R} \mid \G_t^\bullet]= X^{\zeta^2}(t)$, $t \in [0,T]$, and it follows that
$$ \int_0^T \psi^2_u \, dN^{\bullet}_u \ = \ X^{\zeta^2}(t) - \int_0^t \pi^2_u \sigma_u \, dB^\bullet_u -x  \ = \ X^{\zeta^1}(t) - \int_0^t \pi^1_u\sigma_u \, dB^\bullet_u - x \ = \ \int_0^T \psi^1_u \, dN^{\bullet}_u, $$
so we also have $\psi^1 \equiv \psi^2$. 
\hfill{$\Box$}

\begin{remark} \label{remark def z hat}
The optimization problem even has a solution, if we only assume (NA1) in the filtration $\GG^{\bullet}$ since the martingale representation theorem for $\GG^W$ holds by Assumption \ref{ass 2} (b) and for $\GG^S$ it still holds under the weaker no arbitrage assumption (cf.\ \cite[Proposition 3.10.]{Fontana2013}). 
So, the process $Z^\bullet$ may fail to be a true martingale and we cannot define $P^\bullet$ as a probability measure and the process $B^\bullet$ is only a drifted $P$-Brownian motion. 
Nevertheless, we obtain a similar result as in Theorem \ref{thm g w}: 
The unique optimal risk-minimizing trading strategy $\hat{\zeta}=(\hat{\pi},\hat{\psi}) \in \Pi(\mathbb{G}^{\bullet},x)$ and the corresponding wealth process $X^{\hat{\zeta}}$ satisfy
$$ X^{\hat{\zeta}}(t) \ = \ x + \int_{0}^{t} \hat{\pi}_{u}\sigma(u,\tilde{S}_u) \, dB^\bullet_{u} + \int_0^t \hat{\psi}_u \, dN^{\bullet}_u \ = \ (Z^\bullet)_t^{-1} \EE\left[ Z^\bullet_T \hat{R} \, | \, \mathcal{G}^\bullet_{t}\right], \quad t \in [0,T]. $$
\end{remark}

Again, let us end this subsection by giving an example for the same utility and loss functions as in Example \ref{example1}. 

\begin{example}
Let $U(k)=\ln k$ and $L(k)= - \frac3k$ be given. 
Now, the optimal terminal value for Problem \ref{problem1} is given by
$$ \hat{R} \ = \  \frac{1+\sqrt{1 + 12 \lambda^* \hat{y} Z^\bullet_T}}{2\hat{y} Z^\bullet_T}, $$
where $\hat{y}$ and $\lambda^*$ are such that equations \eqref{eq y lambda2} are satisfied. \\
For the optimal wealth process $X^{\hat{\zeta}}$ it holds that
$$ X^{\hat{\zeta}}(t) \ = \ x + \int_0^t \hat{\pi}_u \, d\tilde{S}_u + \int_0^T \hat{\psi}_u \, dN^{\bullet}_u \ = \ x + \int_0^t \hat{\pi}_u \sigma_u \, dB^\bullet_u + \int_0^T \hat{\psi}_u \, dN^{\bullet}_u \ = \ \EE \left[\left.\frac{1+\sqrt{1 + 12 \lambda^* \hat{y} Z^\bullet_T}}{2\hat{y} Z^\bullet_T} \, \right| \, \G^{\bullet}_t \right]. $$
Moreover, the output for the investor is given by
$$ u(\GG^{\bullet},x) \ = \ \EE[U(\hat{R})] \ = \ \EE \left[\ln \left(\frac{1+\sqrt{1 + 12 \lambda^* \hat{y} Z^\bullet_T}}{2\hat{y} Z^\bullet_T}\right) \right]. $$
\end{example}

\subsection{The price process filtration} \label{subsec f s}

Let us now consider the filtration $\mathbb{F}^S:=(\mathcal{F}^S_t)_{t \in [0,T]}$, where we recall that $\mathcal{F}^S_t$ represents the knowledge of the price process $S$ of the risky asset up to time $t$, but where the random time is not necessarily an $\mathbb{F}^S$-stopping time. 
The filtration is motivated by financial markets where an investor has no additional information, e.\,g.\ when the market parameters will change. \\

Let us first sate the representation for the price process $\tilde{S}$ w.\,r.\,t.\ this filtration. 

\begin{lemma} \label{lemma fs decomposition of s} {\em (cf.\ \cite[Lemma 4.2.]{Fontana2013})}
Suppose that Assumptions \ref{ass 1} and \ref{ass 2} (a) hold true. 
There exist an $\mathbb{F}^S$-Brownian motion $\bar{W}=(\bar{W}_t)_{t \in [0,T]}$, as well as a square-integrable, $\mathbb{F}^S$-predictable process $\bar{\mu}=(\bar{\mu}_t)_{t \in [0,T]}$, such that the $\mathbb{F}^S$-semimartingale decomposition of $\tilde{S}$ is written as
\begin{eqnarray*} 
\tilde{S}_t & = & \int_0^t \bar{\mu}_u \, du + \int_0^t \sigma_u \, d\bar{W}_u, \qquad \mbox{for all } t \in [0,T], 
\end{eqnarray*}
where $\bar{\mu}$ is defined as
\begin{eqnarray*} 
\bar{\mu}_t & := & ^p({\bf 1}_{[0,\tau]})_t \mu^1(t,\tilde{S}(t)) + \, ^p({\bf 1}_{(\tau,T]})_t \mu^2(t,\tilde{S}(t)) + \, ^p({\bf 1}_{[0,\tau]} \theta)_t \sigma^1(t,\tilde{S}(t)) + \, ^p({\bf 1}_{(\tau,T]} \theta)_t \sigma^2(t,\tilde{S}(t)),
\end{eqnarray*}
where $^p$ denotes the $\mathbb{F}^S$-predictable projection, and $\sigma=(\sigma_t)_{t \in [0,T]}$ is given in \eqref{eq sigma general}.
\end{lemma}

As already mentioned in Remarks \ref{remark def z} and \ref{remark def z hat}, we consider a weaker condition on no arbitrage: \\

\textbf{Assumption.} 
It holds (NA1) in the filtration $\FF^S$. \\
This is equivalent to $\int_0^T (\bar{\mu}_u/\sigma_u)^2du <+\infty$, $P$-a.s., cf.\ \cite[Proposition 4.3.]{Fontana2013}. \\

Let us define a process $\bar{Z}=(\bar{Z}_{t})_{t \in [0,T]}$ by
$$ \bar{Z}_t \ := \ \exp \left\{ -\int_{0}^{t} \frac{\bar{\mu}_{u}}{\sigma_u} \, d\bar{W}_{u} - \frac12 \int_{0}^{t} \left(\frac{\bar{\mu}_{u}}{\sigma_u}\right)^{2} \, du \right\}, \qquad \mbox{for all } t \in [0,T]. $$
Moreover, define a drifted $\FF^S$-Brownian motion $\bar{B}=(\bar{B}_{t})_{t \in [0,T]}$ by 
$$ \bar{B}_{t} \ = \ \bar{W}_{t} + \int_{0}^{t} \frac{\bar{\mu}_{u}}{\sigma_u} \, du, \qquad 0 \leq t \leq T, $$
which is by assumption of (NA1) well-defined. \\

Let us distinguish the following three cases:
\begin{enumerate}
\item[(a)] Identical volatilities:
$$ \sigma^1(t,x) \ = \sigma^2(t,x) \ =: \ \sigma(t,x), $$
for all $(t,x) \in [0,T] \times \RR$.
\item[(b)] Distinct volatility functions:
$$ \sigma^1(t,x) \ \neq \ \sigma^2(t,x), $$
for any $(t,x) \in [0,T] \times \RR$.
\item[(c)] The volatilities differ on an open set:
$$ \mathcal{O} \ := \ \left\{(t,x) \in [0,T] \times \RR \ : \ \sigma^1(t,x) \neq \sigma^2(t,x) \right\}, $$
which is neither the empty set nor the whole space $[0,T] \times \RR$.
\end{enumerate}
We examine the three cases in the next subsections.

\subsubsection{Identical volatilities} \label{subsec identical volatilities}

{\bf Assumption:} $ \sigma^1(t,x) = \sigma^2(t,x) =: \sigma(t,x)$, for all $(t,x) \in [0,T] \times \RR$. \\

In this case, the canonical decomposition of $\tilde{S}$ in the filtration $\mathbb{F}^S$ is given by (see Lemma \ref{lemma fs decomposition of s}):
$$ \tilde{S}_t \ = \ \int_0^t \bar{\mu}_u \, du + \int_0^t \sigma(u,\tilde{S}_u) \, d\bar{W}_u, \qquad \mbox{for all } t \in [0,T]. $$
The continuous process $\sigma(\cdot,\tilde{S}_\cdot)$ is $\mathbb{F}^S$-adapted and never hits zero, so it is $\mathbb{F}^S$-predictable. \\


Similar to Definition \ref{def trading strategy g w tau}, we denote by $\Pi(\mathbb{F}^{S},x)$  and $\X(\mathbb{F}^{S},x)$ the sets of {\em admissible trading strategies} and of corresponding wealth processes, respectively, where for $\pi \in  \Pi(\mathbb{F}^{S},x)$ we have
\begin{equation} \label{eq wealth process v in f s}
X^{\pi}(t) \ = \ x + \int_{0}^{t} \pi_{u} \, d\tilde{S}_{u} \ = \ x + \int_{0}^{t} \pi_{u}\bar{\mu}_{u}\, du + \int_{0}^{t} \pi_{u} \sigma(u,\tilde{S}_u) \, d\bar{W}_{u}, \qquad 0 \leq t \leq T.
\end{equation}
where the initial capital $x \in \RR$ is assumed to be greater than zero. 
By the assumptions of (NA1) and $\int_{0}^{T} (\pi_{u}\sigma(u,\tilde{S}_u))^{2} du < +\infty$, $P$-a.s., equation \eqref{eq wealth process v in f s} is well-defined, since by the Cauchy-Schwarz inequality it holds that $\int_0^T \pi_t \bar{\mu}_t \, dt < +\infty$, $P$-a.s. \\

We have the main result for this subsection:

\begin{theorem} \label{thm f s}
Suppose Assumptions \ref{ass 1} and \ref{ass 2} (a) as well as (NA1) hold true. 
Moreover, let $\EE[\bar{Z}_T \cdot \tilde{I}_\lambda(y\bar{Z}_T)]<+\infty$ for all $y \in (0,+\infty)$. 
Then the optimal terminal wealth $\bar{R}$ for Problem \ref{problem1} is given by
$$ \bar{R} \ = \ \tilde{I}_{\lambda^*}\left( \hat{y}\bar{Z}_T \right), $$
where $\hat{y} \in (0,+\infty)$ and $\lambda^* \geq 0$ are such that
\begin{equation} \label{eq existence y lambda4}
\left\{ \begin{array} {rcl}  \EE\left[\bar{Z}_T \cdot \tilde{I}_{\lambda^*}(\hat{y}\bar{Z}_T)\right] & =& x, \\ 
\EE\left[L(-\bar{R})\right] & = & \varepsilon. \end{array} \right. 
\end{equation}
The unique optimal trading strategy $\bar{\pi} \in \Pi(\mathbb{F}^{S},x)$ and the corresponding wealth process $X^{\bar{\pi}}$ satisfy:
$$ X^{\bar{\pi}}(t) \ = \ x + \int_{0}^{t} \bar{\pi}_{u}\sigma(u,\tilde{S}_u) \, d\bar{B}_{u} \ = \ \bar{Z}_t^{-1} \cdot \EE\left[ \bar{Z}_T\bar{R} \, | \, \mathcal{F}^{S}_{t}\right]. $$
\end{theorem}

{\em Proof.} 
By the definition of the processes $\bar{Z}$ and $\bar{B}$, the process $\tilde{S}$ is of the form
\begin{eqnarray*}
\tilde{S}_{t} & = & \int_{0}^{t} \bar{\mu}_{u} \, du + \int_{0}^{t} \sigma(u,\tilde{S}_u) \, d\bar{W}_{u} \ = \ \int_{0}^{t} \sigma(u,\tilde{S}_u) \, d\bar{B}_{u}, \qquad \mbox{for all } t \in [0,T],
\end{eqnarray*}
and for $\pi \in \Pi(\FF^S,x)$ the wealth process $X^{\pi}$ admits the representation
\begin{eqnarray*}
X^{\pi}(t) & = & x + \int_{0}^{t} \pi_{u} \, d\tilde{S}_{u} \ = \ x + \int_{0}^{t} \pi_{u} \sigma(u,\tilde{S}_u) \, d\bar{B}_{u}, \qquad \mbox{for all } t \in [0,T].
\end{eqnarray*}
Therefore, $X^{\pi}$ is $\mathbb{F}^{S}$-adapted. \\

Now, it holds that every process $\bar{M}=(M_t)_{t \in [0,T]}$ such that the process $\bar{Z}\bar{M}=(\bar{Z}_t\bar{M}_t)_{t \in [0,T]}$ is a martingale w.\,r.\,t.\ the filtration $\FF^S$ has the representation 
\begin{equation} \label{eq mrt for tilde p in f s 2} 
\bar{M}_{t} \ = \ \bar{M}_{0} + \int_{0}^{t} \pi_{u} \sigma(u,S_u) \, d\bar{B}_{u}, \qquad 0 \leq t \leq T.
\end{equation}
for some $\pi \in \Pi(\FF^{S},x)$, by the same arguments as in Lemma \ref{mrt g w tau}, where we use the martingale representation property (cf.\ \cite[Proposition 4.6.]{Fontana2013}): 
Every $\mathbb{F}^S$-local martingale $\bar{L}=(\bar{L}_t)_{t \in [0,T]}$ admits a representation of the form
$$ \bar{L}_t \ = \ \bar{L}_0 + \int_0^t \bar{\varphi}_u \, d\bar{W}_u, \qquad \mbox{ for all } t \in [0,T], $$
where $\bar{\varphi}=(\bar{\varphi}_t)_{t \in [0,T]}$ is some $\mathbb{F}^S$-predictable process with $\int_0^T \bar{\varphi}_t^2 dt <+\infty$, $P$-a.s. \\

Next we show that for $\bar{R}$ there exists a trading strategy $\bar{\pi} \in \Pi(\mathbb{F}^{S},x)$ such that $X^{\bar{\pi}}(t) = \bar{Z}^{-1}_t \cdot \EE\left[\bar{Z}_T \bar{R} \, | \, \mathcal{F}^{S}_{t}\right]$. 
Define $\bar{M}(t) = \bar{Z}^{-1}_t\EE\left[\bar{Z}_T \bar{R} \, | \, \mathcal{F}^{S}_{t}\right]$, $0 \leq t \leq T$. 
Because for $0 \leq s \leq t$ it holds:
\begin{eqnarray*}
\EE \left[\bar{Z}_t \bar{M}(t) \, | \, \mathcal{F}^{S}_{s}\right] & = & \EE\left[\bar{Z}_t \bar{Z}^{-1}_t \cdot \EE\left[\bar{Z}_T \bar{R} \, | \, \mathcal{F}^{S}_{t}\right] \, | \, \mathcal{F}^{S}_{s}\right] \\
& = & \bar{Z}_s \bar{Z}^{-1}_s \cdot \EE\left[\bar{Z}_T \bar{R} \, | \, \mathcal{F}^{S}_{s}\right]  \qquad \qquad = \ \bar{Z}_s \bar{M}(s),
\end{eqnarray*}
$\bar{Z} \bar{M}$ is a martingale w.\,r.\,t.\ the filtration $\mathbb{F}^{S}$. 
By \eqref{eq mrt for tilde p in f s 2}, we obtain:
$$ \bar{M}(t) \ = \ \bar{M}(0) + \int_{0}^{t} \bar{\pi}_{u} \sigma_{u} \, d\bar{B}_{u}, \qquad 0 \leq t \leq T, $$
with $\bar{M}(0)=\bar{Z}_0\EE\left[\bar{Z}_T \bar{R} \, | \, \mathcal{F}^{S}_{0}\right] = x$. 
So, it holds that $X^{\bar{\pi}}(t) = \bar{M}(t) = \bar{Z}_t\EE\left[\bar{Z}_T \bar{R} \, | \, \mathcal{F}^{S}_{t}\right]\geq 0$, since $\bar{Z}_T\bar{R}$ is non-negative. 
So, $\pi$ is indeed in $\Pi(\FF^{S},x)$. \\

The existence of $\hat{y} \in (0,+\infty)$ and $\lambda^* \geq 0$ such that equation \eqref{eq existence y lambda4} holds is proven in Appendix \ref{app lagrangian}. \\
For the rest of the proof, just follow the arguments in Theorem \ref{theorem optimal trading strategy g w tau}. 
\hfill{$\Box$} \\

In this setting, the financial market, in which the asset $S$ can be traded w.\,r.\,t.\ its own generated filtration $\mathbb{F}^S$, is complete, in the sense that all bounded $\mathcal{F}^S_T$-measurable contingent claims can be perfectly replicated by an admissible trading strategy -- even we only claim (NA1) instead of (NFLVR), cf.\ \cite[Corollary 4.7.]{Fontana2013}.

\subsubsection{Totally distinct volatilities}

{\bf Assumption:} $ \sigma^1(t,x) \ \neq \ \sigma^2(t,x)$, for any $(t,x) \in [0,T] \times \RR$. \\

This condition implies that the volatility of $\tilde{S}$ is different on the whole interval $(\tau,T]$ after the stopping time $\tau$. 
In this case, all results of Subsection \ref{subsec g w} are valid for the filtration $\FF^S$, since we have the following result.

\begin{proposition} {\em (cf.\ \cite[Proposition 4.4.]{Fontana2013})} 
Suppose Assumptions \ref{ass 1} and \ref{ass 2} (a) hold true. 
Then, $\tau$ is an $\mathbb{F}^S$-stopping time and the filtrations $\mathbb{F}^S$ and $\mathbb{G}^S$ coincide.
\end{proposition}


\subsubsection{Semi-identical volatilities}

{\bf Assumption:} $ \mathcal{O} := \left\{(t,x) \in [0,T] \times \RR \ : \ \sigma^1(t,x) \neq \sigma^2(t,x) \right\} \not\subseteq \{\emptyset, [0,T] \times \RR\}$ and $\sigma^1(0,0) \neq \sigma^2(0,0)$. \\

Define a non-decreasing sequence $(\rho_k)_{k \in \NN}$ of random variables, defined iteratively by:
\begin{eqnarray*}
\rho_0 & = & 0, \\
\rho_{2k-1} & = & \inf \{\rho_{2k-2} < t \leq T \, : \, (t,S_t) \notin \mathcal{O} \} \vee T, \\
\rho_{2k} & = & \inf \{\rho_{2k-1} < t \leq T \, : \, (t,S_t) \in \mathcal{O} \} \vee T.
\end{eqnarray*}

\begin{lemma} {\em (cf.\ \cite{Fontana2013})}
For any $k \in \NN$, the random variable $\rho_k$ is an $\mathbb{F}^S$-stopping time.
\end{lemma}

The members of the sequence $(\rho_k)_{k \in \NN}$ are chosen in the way, that the two volatilities $\sigma^1(t,S_t)$ and $\sigma^2(t,S_t)$ are equal on the union of closed intervals 
$$ I_2 \ := \ \bigcup_{k=1}^\infty [\rho_{2k-1},\rho_{2k}] $$ 
and not equal on the union of open intervals 
$$ I_1 \ :=\ \bigcup_{k=1}^\infty (\rho_{2k-2},\rho_{2k-1}), $$ 
because $\mathcal{O}$ is an open set. \\

Similar to Definition \ref{def trading strategy g w}, we denote by $\Pi(\FF^S,x)$ and $\X(\FF^S,x)$ the sets of {\em admissible trading strategies} and of corresponding wealth processes, respectively, where for $\zeta=(\pi,\psi) \in \Pi(\FF^S,x)$ we have for $0 \leq t\leq T$
\begin{equation} \label{eq wealth process v in f s3}
X^{\zeta}(t) \ = \ x + \int_{0}^{t} \pi_{u} \, d\tilde{S}_{u} + \int_0^t \psi_u \, dN^{S}_u \ = \ x + \int_{0}^{t} \pi_{u}\bar{\mu}_{u}\, du + \int_{0}^{t} \pi_{u} \sigma_{u} \, d\bar{W}_{u} + \int_0^t \psi_u \, dN^{S}_u, 
\end{equation}
where the initial capital $x \in \RR$ is assumed to be greater than zero and $N^{S}$ is defined as in Lemma \ref{lemma representation s on g s} (ii). 
By the assumptions of (NA1) and $\int_{0}^{T} (\pi_{u}\sigma_{u})^{2} du < +\infty$, $P$-a.s., \eqref{eq wealth process v in f s3} is well-defined, since by the Cauchy-Schwarz inequality it holds that $\int_0^T \pi_t \bar{\mu}_t \, dt < +\infty$, $P$-a.s. \\
Moreover, we will consider {\em risk-minimizing} trading strategies as defined in Definition \ref{def risk minimizing strategy}. \\


The main result of this subsection is the following: 

\begin{theorem} \label{thm f s3}
Suppose Assumptions \ref{ass 1} and \ref{ass 2} (a) as well as (NA1) hold true. 
Moreover, let $\EE[\bar{Z}_T \cdot \tilde{I}_\lambda(y\bar{Z}_T)]<+\infty$ for all $y \in (0,+\infty)$. 
Then the optimal terminal wealth $\hat{R}$ for Problem \ref{problem1} is given by
$$ \hat{R} \ = \ \tilde{I}_{\lambda^*}\left( \hat{y}\bar{Z}_T \right), $$
where $\hat{y} \in (0,+\infty)$ and $\lambda^* \geq 0$ are such that
\begin{equation} \label{eq y lambda4} 
\left\{ \begin{array} {rcl}  \EE\left[\bar{Z}_T \cdot \tilde{I}_{\lambda^*}(\hat{y}\bar{Z}_T)\right] & = & x, \\ 
\EE\left[L(-\hat{R})\right] & = & \varepsilon. \end{array} \right. 
\end{equation}
The unique optimal risk-minimizing trading strategy $\hat{\zeta}=(\hat{\pi},\hat{\psi}) \in \Pi(\mathbb{F}^{S},x)$ and the corresponding wealth process $X^{\hat{\zeta}}$ satisfy:
$$ X^{\hat{\zeta}}(t) \ = \ x + \int_{0}^{t} \hat{\pi}_{u}\sigma_u \, d\bar{B}_{u} +\sum_{k=1}^\infty \int_{\rho_{2k-2} \vee t}^{\rho_{2k-1} \vee t} \hat{\psi}_u \, dN^{S}_u \ = \ \bar{Z}_t^{-1} \EE\left[ \bar{Z}_T \hat{R} \, | \, \mathcal{F}^S_{t}\right]. $$
\end{theorem}

For the proof of the theorem let us first state the martingale representation theorem which is a consequence of the corresponding martingale representation theorems for $\GG^S$ and $\FF^S$ when the volatilities are totally identical. 

\begin{lemma} \label{lemma mrt for hat p in f s3}
Under the same assumptions as in Theorem \ref{thm f s3}, it holds that every process $\hat{L}=(\hat{L}_{t})_{t \in [0,T]}$ such that the process $\bar{Z}\hat{L}=(\bar{Z}_t\hat{L}_t)_{t \in [0,T]}$ is a martingale w.\,r.\,t.\ the filtration $\mathbb{F}^{S}$ is of the form
\begin{equation} \label{eq mrt for hat p in f s 3} 
\hat{L}_{t} \ = \ \hat{L}_0 + \int_0^t \pi_u \sigma_u \, d\bar{B}_u + \sum_{k=1}^\infty \int_{\rho_{2k-2} \vee t}^{\rho_{2k-1} \vee t} \psi_u \, dN^{S}_u, \qquad 0 \leq t \leq T,
\end{equation}
where $\pi=(\pi_t)_{t \in [0,T]}$ and $\psi=(\psi_t)_{t \in [0,T]}$ are $\mathbb{F}^S$-predictable stochastic processes satisfying $\int_0^T (\pi_u \sigma_u)^2 du < +\infty$, $P$-a.s., and $\int_0^T {|\hat{\psi}_t| |dA^{S}_t|} dt < +\infty$, $P$-a.s.
\end{lemma}

\textit{Proof.}
Because the martingale representation property can also be defined on local sets, any $\mathbb{F}^S$-local martingale has the representation given in Lemma \ref{lemma mrt g w} on the sets $(\rho_{2k-2},\rho_{2k-1})$, $k \in \NN$, and the representation given in \eqref{eq mrt for tilde p in f s 2} on the sets $[\rho_{2k-1},\rho_{2k-1}]$, $k \in \NN$. 
\hfill{$\Box$} \\

We now give the proof of the main result of this subsection. \\

{\bf Proof of Theorem \ref{thm f s3}.} 
The existence of $\hat{y} \in (0,+\infty)$ and $\lambda^* \geq 0$ such that equations \eqref{eq y lambda4} hold is proven in Appendix \ref{app lagrangian}. \\
Let us show that there exists a trading strategy $\hat{\zeta}=(\hat{\pi},\hat{\psi}) \in \Pi(\mathbb{F}^S,x)$, such that for the corresponding wealth process $X^{\hat{\zeta}}$ it holds:
$$ X^{\hat{\zeta}}(t) \ = \ \bar{Z}_t^{-1} \EE\left[ \bar{Z}_T \hat{R} \, | \, \mathcal{F}^{S}_{t}\right]. $$
Define $M(t) = \bar{Z}_t^{-1} \EE\left[\bar{Z}_T \hat{R} \, | \, \mathcal{F}^{S}_{t}\right]$, $0 \leq t \leq T$. 
By the same arguments as in the proof of Theorem \ref{thm f s} it holds that $\bar{Z}M$ is a $P$-martingale w.\,r.\,t.\ the filtration $\mathbb{F}^{S}$. 
By Lemma \ref{lemma mrt for hat p in f s3}, we obtain that there exists $\hat{\zeta}=(\hat{\pi},\hat{\psi}) \in \Pi(\mathbb{F}^S,x)$ such that
$$ M(t) \ = \ M(0) + \int_{0}^{t} \pi_{u} \sigma_{u} \, d\bar{B}_{u} + \sum_{k=1}^\infty \int_{\rho_{2k-2} \vee t}^{\rho_{2k-1} \vee t} \hat{\psi}_u \, dN^{S}_u, \qquad 0 \leq t \leq T, $$
with $M(0)=\EE[\bar{Z}_T \hat{R}] = x$. 
So, it holds that $X^{\hat{\zeta}}(t) = M(t) = \bar{Z}_t^{-1} \EE\left[\bar{Z}_T \hat{R} \, | \, \mathcal{F}^{S}_{t}\right]$. \\

The optimality and the uniqueness of $\hat{\zeta}$ follow by similar arguments as in the proof of Theorem \ref{thm g w}.
\hfill{$\Box$} \\

Again, let us end this subsection by giving an example for the same utility and loss functions as in Example \ref{example1}. 

\begin{example}
Let $U(k)=\ln k$ and $L(k)= - \frac3k$ be given. 
Now, the optimal terminal value for Problem \ref{problem1} is given by
$$ \hat{R} \ = \  \frac{1+\sqrt{1 + 12 \lambda^* \hat{y} \bar{Z}_T}}{2\hat{y} \bar{Z}_T}, $$
where $\hat{y}$ and $\lambda^*$ are such that equations \eqref{eq y lambda4} are satisfied. \\
For the optimal wealth process $X^{\hat{\zeta}}$ it holds that
$$ X^{\hat{\zeta}}(t) \ = \ 
x + \int_0^t \hat{\pi}_u \sigma_u \, d\bar{B}_u + \sum_{k=1}^\infty \int_{\rho_{2k-2}}^{\rho_{2k-1}} \hat{\psi}_u \, dN^{S}_u \ = \ \EE \left[\left.\frac{1+\sqrt{1 + 12 \lambda^* \hat{y} \bar{Z}_T}}{2\hat{y} \bar{Z}_T} \, \right| \, \F^S_t \right]. $$
Moreover, the output for the investor is given by
$$ u(\FF^S,x) \ = \ \EE[U(\hat{R})] \ = \ \EE \left[\ln \left(\frac{1+\sqrt{1 + 12 \lambda^* \hat{y} \bar{Z}_T}}{2\hat{y} \bar{Z}_T}\right) \right]. $$
\end{example}

\subsection{Utility indifference value} \label{subsec uiv}

Let us now evaluate how big the monetary advantage from additional information is. 
For that, let us define the utility indifference value which is mentioned by e.\,g.\ \cite{Amendinger2003}.

\begin{definition}[Utility indifference value]
Let $\FF=(\F_t)_{t \in [0,T]}$ and $\GG=(\G_t)_{t \in [0,T]}$ two of the considered filtrations such that $\FF \subseteq \GG$. 
The {\em utility indifference value} of additional information $\GG$ is defined as a solution $c=c(x)$ of the equation
\begin{equation} \label{eq def uiv}
u(\FF,x) \ = \ u(\GG,x-c). 
\end{equation}
\end{definition}

The utility indifference value describes the maximal value the investor is willing to pay for the additional information such that he is indifferent which information he uses for optimizing his utility from terminal wealth. 
If $u(\GG,\cdot)$ and $u(\FF,\cdot)$ are strictly increasing, continuous and finite as well as there exists $y > 0$ such that $u(\FF,x)>u(\GG,y)$ then there exists a solution. 
Moreover, since $\Pi(\FF,x) \subseteq \Pi(\GG,x)$, it holds that $u(\FF,x) \leq u(\GG,x)$ for all $x$, so $c$ is non-negative. 
To obtain some explicit results we have to assume that the combined utility and loss function has a common type, i.\,e.\ it is of power or logarithmic type.
Therefore, let us consider $U(k)=- \frac1k + 1$ and $L(k)= - \frac3k$. 
Then all properties of Definitions \ref{def utility function} and \ref{def loss function} are satisfied. 
Let us assume that $\varepsilon_{\min} \leq \varepsilon \leq \varepsilon_{\max}$ for the corresponding filtrations. 

\begin{theorem}
Let $\FF,\GG \in \{\widetilde{\GG}^{W},\widetilde{\GG}^{S},\GG^W,\GG^S,\FF^S\}$ such that $\FF \subseteq \GG$ and $Z^\FF$, $Z^\GG$ the corresponding densities. 
Moreover, let Assumption \ref{ass 1} as well as all assumptions from the corresponding subsections hold true. 
Then the utility indifference value is given by
\begin{equation} \label{eq utility indifference price} 
c(x) \ = \ \left(1 - \left(\frac{\EE\left[\sqrt{Z^\GG_T} \mid \G_0\right]}{\EE\left[\sqrt{Z^{\FF}_T} \mid \F_0\right]}\right)^2\right) \cdot x. 
\end{equation}
\end{theorem}

{\em Proof.} 
We have $\tilde{U}_\lambda(k)=-\frac1k + 1 - \frac{3\lambda}k$ and $\tilde{I}_\lambda(k)=\sqrt{\frac{1+3\lambda}{k}}$. 
Let $\HH \in \{\FF,\GG\}$ and let $Z^\HH$ be the corresponding density process defined in the several subsections. 
Then the optimal terminal wealth is given by
$$ \hat{R} \ = \ \sqrt{\frac{1+3\lambda^*}{\hat{y}Z_T^\HH}}, $$
where it holds that 
$$\hat{y} \ = \ \left(\frac{\EE[\sqrt{Z^\HH_T (1+3 \lambda^*)}]}{x}\right)^2. $$
Moreover, the value process $u$ admits the representation
\begin{equation} \label{eq optimal u} 
u(\HH,x) \ = \ 1 - \frac{1}x \cdot \left(\EE\left[\sqrt{Z^\HH_T} \mid \Hcal_0\right]\right)^2.
\end{equation}
We just replace $\FF$ and $\GG$ in \eqref{eq optimal u} and calculate the corresponding $c$. 
\hfill{$\Box$} \\

We see that the utility indifference price only depends on the density processes $Z^\FF$ and $Z^\GG$. 
By the positivity condition of \eqref{eq utility indifference price} it holds that $\EE[\sqrt{Z^\FF_T} \mid \F_0] \geq \EE[\sqrt{Z^\GG_T} \mid \G_0]$ where the difference of the two conditional expectations the larger the more information is contained in $\GG$ compared to $\FF$. 
Note that if $\FF$ or $\GG$ are initially enlarged filtrations, then the utility indifference price is a random variable.

\section{Conclusion} \label{sec conclusion}

In this paper we considered an expected utility maximization problem under a risk constraint and with incomplete information about the underlying Brownian motion and a random change point. 
This is modeled by different filtrations in which the random time is a stopping time or is deterministic. 
The agent's preference is measured by a general utility function and the risk is measured by a utility-based shortfall risk. 
For five different filtration we gave the optimal solution under mild assumptions by using martingale methods. 
Moreover, we calculated the utility indifference value for a special utility and loss function which measures the gain of additional information for the investor. \\
In contrast to other authors we only considered one stock driven by one Brownian motion. 
Future research could extend it to a multi-dimensional model, even with more than one change point. 
Moreover, instead of considering an maximization problem for the optimal terminal wealth one can add a consumption process to the model, similar to the setting in \cite{Lakner1995}. \\

One interesting question for further research is under which assumption one can give an explicit formula for the optimal trading strategies. 
Moreover, the expected utility and the expected loss are modeled w.\,r.\,t.\ the same filtration. 
One can ask for optimal solutions when the maximizing agent and the risk-regulator have different information modeled by different filtrations. 
This question also deals with filtering techniques and is part of ongoing research.

\begin{appendix}

\section{Existence of Lagrange multipliers} \label{app lagrangian}

In this section we prove the existence of the two Lagrange multipliers $y$ and $\lambda$ such that the budget constraint as well as the the risk constraint are satisfied. 
The proof is valid for all filtrations considered in this paper, we will write the general filtration $\FF \in \{\widetilde{\GG}^{W},\widetilde{\GG}^{S},\GG^W,\GG^S,\FF^S\}$ with corresponding $\F_t \in \{\widetilde{\G}^{W}_t,\widetilde{\G}^{S}_t,\G^W_t,\G^S_t,\F^S_t\}$, $t \in [0,T]$, and consider the corresponding general density process $Z \in \{\widetilde{Z}^{W},\widetilde{Z}^{S},Z^{W},Z^S,\bar{Z}\}$. \\

We extend the arguments of \cite[Lemma 6.1]{Gundel2008} for non-trivial $\sigma$-fields $\widetilde{\G}^{W}_0$ and $\widetilde{\G}^{S}_0$. 
Therefore, for these filtrations the multipliers are no longer real values, but $\widetilde{\G}^{W}_0$- and $\widetilde{\G}^{S}_0$-measurable random variables. \\

Let $U$ and $L$ be the utility and loss function as defined in Definition \ref{def utility function} and Definition \ref{def loss function}, respectively. 
Let us denote the inverse of the first derivative of $L$ by $H:=(L')^{-1}$. 
We define 
\begin{eqnarray*}
g_{\lambda,y}(x) & := & U(x) - \lambda L(-x) - y x, \qquad x \in (0,+\infty). 
\end{eqnarray*}

Then it holds:

\begin{lemma} {\em (cf.\ \cite[Lemma A.1.]{Gundel2008})} \label{prop x star}
\begin{enumerate}
\item[(i)] 
$g_{\lambda,y}$ is strictly concave and continuous on $(0,+\infty)$. 
\item[(ii)] 
The maximizer of $g_{\lambda,y}$ is given by $x^*(\lambda,y)$ which is the unique solution of the equation $U'(x)+\lambda L'(-x)=y$.
\item[(iii)] 
The function $x^*:[0,\infty)\times(0,\infty) \to (0,\infty)$ is continuous. 
\item[(iv)] 
$x^*(\lambda,y)$ is decreasing in $y$ for $\lambda$ fixed, and increasing in $\lambda$ for $y$ fixed.
\item[(v)] 
We have $\lim_{y \to \infty} x^*(\lambda,y)=0$ and $\lim_{y \to 0} x^*(\lambda,y)=+\infty$ for fixed $\lambda$.
\item[(vi)] 
For $\alpha \geq 1$ it holds that $x^*(\alpha \lambda,\alpha y) \leq x^*(\lambda,y)$. 
\item[(vii)] 
The inverse function $H$ of the first derivative of the loss function $L$ is continuous and strictly increasing on $[0,L'(0-)]$.
If $e \in (0,L'(0-))$ and $\mu:=U'(H(e))$, then we have for all $\lambda \geq 0$, that $x^*(0,\mu)=x^*(\lambda,\mu+\lambda e)$. 
\item[(viii)]
Let $\tilde{c}:\RR_+ \to \RR_+$ be decreasing with $\lim_{x \to \infty} \tilde{c}(x)=c > 0$. 
Then it holds that $\lim_{\lambda \to \infty} x^*(\lambda, \tilde{c}(\lambda) \cdot \lambda)=-H(c)$. 
Moreover, $x^*(\lambda,c \lambda)$ converges to $-H(c)$ monotonously from above. 
\item[(ix)] 
It holds that
$$ \sup_{x > 0} \left\{-\lambda L(-x) - y x \right\} \ = \ -\lambda L \left( H \left(\frac{y}{\lambda}\right)\right) + y H \left(\frac{\lambda_2}{\lambda_1}\right). $$
\end{enumerate}
\end{lemma}

Now, we show that the Lagrange multipliers for the budget and the risk constraint exist and that both constraints are satisfied by equality.

\begin{lemma} \label{lemma existence lagrange multipliers}
Let the assumption \eqref{ass u finite} hold true and let the values \eqref{eq epsilon min} and \eqref{eq epsilon max} exist. 
Moreover, let $\EE[Z_T \cdot \tilde{I}_{\lambda}( yZ_T )) \mid \F_{0}] < +\infty$, $P$-a.s. 
Then there exist $\F_0$-measurable random variables $\hat{y} \in (0,+\infty)$ and $\lambda^* \geq 0$ such that for the optimal solution 
$$ \hat{R} \ = \ \tilde{I}_{\lambda^*}\left( \hat{y}Z_T \right), $$
it holds that
$$ \left\{ \begin{array} {rcl}  \EE\left[Z_T \cdot \hat{R} \mid \F_{0}\right] & =& x, \quad P\text{a.s.,}\\ 
\EE\left[L(-\hat{R}) \mid \F_0 \right] & = & \varepsilon, \quad P \text{-a.s.} \end{array} \right. $$
\end{lemma}

We first show that there exists a $\hat{y} \in (0,+\infty)$ such that the budget constraint $\EE[Z_T \cdot \tilde{I}_{\lambda}( \hat{y}Z_T )) \mid \F_0] = x$, $P$-a.s., is satisfied for every $\lambda \geq 0$. \\

For fixed $\lambda \geq 0$ let us define the function $\Hcal_{\omega}:(0,+\infty) \to (0,+\infty]$ by
$$ \Hcal_\omega(y) \ := \ \EE\left[Z_T \cdot \tilde{I}_{\lambda}( yZ_T )) \, \left| \, \F_{0}\right.\right](\omega). $$

Then we have the following result.

\begin{lemma}\label{lemma function h g w tau} {\em (cf.\ \cite[Lemma 5.2]{amendinger})}
If it holds that $\Hcal_\omega$ is finite for $P$-almost all $\omega \in \Omega$, then there exists a $\F_0$-measurable random variable $\hat{y} \in (0,+\infty)$ such that $\Hcal_\omega(\hat{y})=x$, and which is for $P$-a.e.\ $\omega \in \Omega$ uniquely defined.
\end{lemma}

Next, we prove the the existence of the second Lagrange multiplier. This is done in several steps. 

\begin{lemma}
Under the same assumptions as in Lemma \ref{lemma existence lagrange multipliers} 
let $\hat{y}(\lambda)$ be the value such that the budget constraint is satisfied. 
Then the function $\hat{y}(\lambda)/\lambda$ is $P$-a.s.\ decreasing for $\F_0$-measurable $\lambda \in (0,+\infty)$ and in particular it holds that the limit
$$ \lim_{\lambda \to \infty} \frac{\hat{y}(\lambda)}{\lambda} \ \in \ [0,+\infty) $$
exists, $P$-a.s.
\end{lemma}

{\em Proof.} 
Let $0 < \lambda < \mu$ and define $\alpha:=\mu/\lambda > 1$. 
From Lemma \ref{prop x star} (vi) we have that
$$ x^*(\mu,\hat{y}(\mu) Z_T) \ = \ x^*\left(\alpha \lambda, \alpha \lambda \frac{\hat{y}(\mu)}{\mu} Z_T \right) \ \leq \ x^*\left(\lambda,\lambda \frac{\hat{y}(\mu)}{\mu} Z_T \right). $$
From this we get that
$$ x \ = \ \EE\left[Z_T x^*\left(\mu,\hat{y}(\mu) Z_T\right) \mid \F_0\right] \ \leq \ \EE\left[Z_T x^*\left(\lambda,\lambda \frac{\hat{y}(\mu)}{\mu} Z_T\right) \mid \F_0 \right], \qquad P\text{-a.s.} $$
Now, let us suppose $\hat{y}(\lambda)/\lambda$ is not decreasing in $\lambda$ , i.e.\ $\hat{y}(\mu)/\mu > \hat{y}(\lambda)/\lambda$, $P$-a.s. 
By Lemma \ref{prop x star} (iv) let us consider $y \geq \lambda \frac{\hat{y}(\mu)}{\mu} > \lambda$ such that
$$ x \ = \ \EE\left[Z_T x^*\left(\lambda,yZ_T\right) \mid \F_0\right], \qquad P\text{-a.s.} $$
But by Lemma \ref{lemma function h g w tau} the solution $\hat{y}(\lambda)$ is uniquely determined, so we have $y = \lambda \frac{\hat{y}(\mu)}{\mu} = \lambda$, $P$-a.s.; a contradiction. 
Therefore, $\hat{y}(\lambda)/\lambda$ is decreasing. 
\hfill{$\Box$} \\

Define 
$$ \hat{R}(\lambda) \ = \ \tilde{I}_{\lambda}\left( \hat{y}(\lambda) Z_T \right) $$ 
for a non-negative, $\F_0$-measurable random variable $\lambda$. 
This expression is $P$-a.s.\ well-defined, by the uniqueness of $\hat{y}(\lambda)$ for a given $\lambda$. 

\begin{lemma} \label{lemma converging lambda}
Under the same assumptions as in Lemma \ref{lemma existence lagrange multipliers} 
let $(\lambda_n)_{n \in \NN}$ be a sequence of non-negative $\F_0$-measurable random variables converging to $\F_0$-measurable $\lambda \geq 0$, $P$-a.s. 
Then there exists a subsequence $(\lambda_{n_j})_{j \in \NN}$ such that $\hat{R}(\lambda_{n_j})$ converges to $\hat{R}(\lambda)$, $P$-a.s.
\end{lemma}

{\em Proof.} 
Consider the corresponding positive sequence $\hat{y}(\lambda_n)_{n \in \NN}$. 
Then it holds that it is $P$-a.s.\ bounded, since otherwise there would be a subset $N \subset \Omega$ with $P(N)>0$ and a subsequence $(n_k)_{k \in \NN}$ such that subsequence $(\hat{y}(\lambda_{n_k}(\omega)))_{k \in \NN}$ is increasing with $\lim_{k \to \infty} \hat{y}(\lambda_{n_k}(\omega)) = +\infty$ for all $\omega \in N$. 
For $\lambda^*(\omega):=\max_{n \in \NN} \lambda_{n}(\omega)$ it holds due to Lemma \ref{prop x star} (iv)\&(v) that
$$ \hat{R}(\lambda_{n_k}(\omega)) \ \leq \ x^*\left(\lambda^*(\omega),\hat{y}(\lambda_{n_k}(\omega)) Z_T\right) \ \searrow \ 0 \qquad \text{for } k \to \infty, $$
which would imply by the monotone convergence theorem that $x(\omega) \leq 0$ for all $\omega \in N$; a contradiction. \\
Moreover, the sequence $\hat{y}(\lambda_n)_{n \in \NN}$ stays $P$-a.s.\ away from zero, since otherwise there would be a subset $N' \subset \Omega$ with $P(N')>0$ and a subsequence $(n_l)_{l \in \NN}$ such that $(\hat{y}(\lambda_{n_l}(\omega)))_{l \in \NN}$ is decreasing with $\lim_{l \to \infty} \hat{y}(\lambda_{n_l}(\omega)) = 0$ for all $\omega \in N'$. 
For $\lambda_*(\omega):=\min_{n \in \NN} \lambda_{n}(\omega)$ it holds due to Lemma \ref{prop x star} (iv)\&(v) that
$$ \hat{R}(\lambda_{n_l}(\omega)) \ \geq \ x^*\left(\lambda_*(\omega),\hat{y}(\lambda_{n_l}(\omega)) Z_T\right) \ \nearrow \ +\infty \qquad \text{for } l \to \infty, $$
which would imply by the monotone convergence theorem that $x(\omega) = +\infty$ for all $\omega \in N'$; a contradiction. \\
Therefore, for $P$-a.e.\ $\omega \in \Omega$ the sequence $(\lambda_n(\omega))_{n \in \NN}$ converges and there exists a subsequence $(n_j)_{j \in \NN}$ and $y(\omega) \in (0,+\infty)$ such that $\lim_{j \to \infty} \hat{y}(\lambda_{n_j}(\omega))=y(\omega)$. 
With this, we have $\lim_{j \to \infty} \hat{R}(\lambda_{n_j}(\omega)) = x^*(\lambda(\omega),y(\omega) Z_T)$. \\
For $y^*(\omega):=\max_{j \in \NN} \hat{y}(\lambda_{n_j}(\omega))$ and $y_*(\omega):= \min_{j \in \NN} \hat{y}(\lambda_{n_j}(\omega))$ we have the bounds
$$ x^*\left(\lambda_*(\omega),\hat{y}^*(\omega) Z_T\right) \ \leq \ \hat{R}(\lambda_{n_j}(\omega)) \ \leq \ x^*\left(\lambda^*(\omega),\hat{y}_*(\omega) Z_T\right), $$
and by the dominated convergence theorem that
$$ x \ = \ \lim_{j \to \infty} \EE\left[Z_T \cdot \hat{R}(\lambda_{n_j}) \mid \F_{0}\right] \ = \ \EE\left[Z_T \cdot x^*(\lambda,y Z_T) \mid \mathcal{F}_{0}\right], \quad P\text{-a.s.} $$
which completes the proof.
\hfill{$\Box$} \\

Then we conclude the proof by the following 

\begin{lemma}
Let the same assumptions as in Lemma \ref{lemma existence lagrange multipliers} hold true. 
For $P$-a.e.\ $\omega \in \Omega$ the function
$$ k_\omega:[0,+\infty) \ \to \ \RR, \quad \lambda \ \mapsto \ \EE\left[L(-\hat{R}(\lambda)) \mid \F_0\right](\omega) $$
is continuous. 
Moreover, it holds that 
$$ \lim_{\lambda \to 0} k_\omega(\lambda) \ = \ \varepsilon_{\max}(\omega),
\qquad \lim_{\lambda \to \infty} k_\omega(\lambda) \ = \ \varepsilon_{\min}(\omega) $$ 
for $P$-a.e.\ $\omega \in \Omega$, where $\varepsilon_{\max}$ and $\varepsilon_{\min}$ are defined in \eqref{eq epsilon max} and \eqref{eq epsilon min}, respectively. 
\end{lemma}

{\em Proof.} 
We first show that the function $k$ is $P$-a.s.\ continuous:
Let $(\lambda_n)_{n \in \NN}$ be a sequence of positive $\F_0$-measurable random variables converging to a $\F_0$-measurable $\lambda$, $P$-a.s. 
Now, choose a subsequence $(n_j)_{j \in \NN}$ such that 
$\lim_{j \to \infty} \hat{R}(\lambda_{n_j})=\hat{R}(\lambda)$, $P$-a.s. 
Similar to the proof of Lemma \ref{lemma converging lambda} we have for $\lambda_*(\omega):=\min_{j \in \NN} \lambda_{n_j}(\omega)$ and $y^*(\omega):=\max_{j \in \NN} \hat{y}(\lambda_{n_j}(\omega))$ by Lemma \ref{prop x star} (iv)\&(v) that
$$ 0 \ \leq \ L(-\hat{R}(\lambda_{n_j}(\omega))) \ \leq \ L(-x^*(\lambda_*(\omega),y^*(\omega) Z_T)) \qquad \text{for all } j \in \NN \text{ and }P\text{-a.e.\ } \omega \in \Omega $$
such that it follows by the dominated convergence theorem that
$$ \lim_{j \to \infty} k_\omega(\lambda_{n_j}) \ = \ \lim_{j \to \infty} \EE\left[L(-\hat{R}(\lambda_{n_j})) \mid \F_0\right](\omega) \ = \ \EE\left[L(-\hat{R}(\lambda)) \mid \F_0\right](\omega) \ = \ k_\omega(\lambda). $$
Now, let $(\lambda_n)_{n \in \NN}$ be a sequence of positive $\F_0$-measurable random variables converging to $0$, $P$-a.s., then it holds:
$$ \lim_{n \to \infty} k_\omega(\lambda_{n}) \ = \ k_\omega(0) \ = \ \EE\left[L(-\hat{R}(0)) \mid \F_0\right](\omega), $$
where 
$$ \hat{R}(0) \ = \ \tilde{I}_{0}\left( \hat{y}(0)Z_T \right) \ = \ I\left( \hat{y}(0)Z_T \right). $$
This is exactly the result for the utility maximization problem without a risk constraint (cf.\ \cite[Proposition 4.5]{Amendinger2003}), i.e.\ there is no other contingent claim which maximizes the expected utility from terminal wealth. 
Therefore, we have
$$ \lim_{n \to \infty} k_\omega(\lambda_{n}) \ = \ \EE\left[L(-\hat{R}(0)) \mid \F_0\right](\omega) \ = \ \varepsilon_{\max}(\omega), \quad \text{for }P\text{-a.e.\ } \omega \in \Omega. $$
For the other statement, consider the $\F_0$-measurable random variable $c^*:=\lim_{\lambda \to \infty} \hat{y}(\lambda)/\lambda$. 
Obviously, $c^* \geq 0$. 
Now, suppose that $c^*(\omega)=0$ for $\omega \in N \subset \Omega$ satisfying $P(N)>0$. 
Let $\omega \in N$. 
Then we find for every $\varepsilon(\omega) > 0$ a corresponding $\mu(\omega) > 0$ such that for any $\lambda(\omega) \geq \mu(\omega)$ we have by Lemma \ref{prop x star} (iv) that
$$ \hat{R}(\lambda(\omega)) \ \geq \ x^*\left(\lambda(\omega), \lambda(\omega) \varepsilon(\omega) Z_T\right). $$
By this and Lemma \ref{prop x star} (viii), 
it follows that
$$ x(\omega) \ = \ \EE\left[Z_T \hat{R}(\lambda) \mid \F_0 \right](\omega) \ \geq \ \EE\left[Z_T x^*\left(\lambda, \lambda \varepsilon T_T\right) \mid \F_0\right](\omega) \stackrel{\lambda \to \infty}{\searrow} \ -\EE\left[Z_T H\left(\varepsilon Z_T\right) \mid \F_0 \right](\omega), $$
where the right-hand side equals $+\infty$ for $\varepsilon(\omega) \to 0$ by the properties of the loss function $L$; a contradiction. 
Therefore, $c^* > 0$, $P$-a.s. \\
Moreover, if $\lambda > n$ for $n \in \NN$, $P$-a.s., it holds due to Lemma \ref{prop x star} (vi) \& (viii) that
$$ -H\left(\frac{\hat{y}(n)}n \cdot Z_T\right) \ \leq \ x^*\left(\lambda, \lambda \cdot \frac{\hat{y}(n)}n \cdot Z_T\right), \quad P\text{-a.s.} $$
Next, by Lemma \ref{prop x star} (iv) \& (vi) and $\frac{\hat{y}(n)}n \cdot \lambda \geq \hat{y}(\lambda) \geq c^* \lambda$, $P$-a.s., it holds that
$$ x^*\left(\lambda, \lambda \cdot \frac{\hat{y}(n)}n \cdot Z_T\right) \ \leq \ \hat{R}(\lambda) \ \leq \ x^*\left(\lambda, c^* \lambda \cdot Z_T\right) \ \leq \ x^*\left(1, c^* \cdot Z_T\right), \quad P\text{-a.s.} $$
By applying the dominated convergence theorem and Lemma \ref{prop x star} (viii), we obtain that
$$ x \ = \ \EE\left[Z_T \hat{R}(\lambda) \mid \F_0 \right] \ \stackrel{\lambda \to \infty}{\longrightarrow} \ \EE \left[-H(c^* Z_T) \mid \F_0 \right], \qquad P\text{-a.s.} $$
We notice, that therefore 
$$ \lim_{\lambda \to \infty} \hat{R}(\lambda) \ = \ -H(c^* Z_T), \qquad P\text{-a.s.,} $$
and $c^*$ is a solution to the equation 
$$ x \ = \ \EE\left[Z_T H(c Z_T) \right] $$
for $P$-a.e.\ $\omega \in \Omega$. 
Moreover by Lemma \ref{prop x star} (viii), we have that
$$ \lim_{\lambda \to \infty} k_\omega(\lambda) \ = \ \lim_{\lambda \to \infty} \EE\left[L(-\hat{R}(\lambda)) \mid \F_0\right](\omega) \ = \ \EE\left[L(H(c^* \cdot Z_T)) \mid \F_0 \right](\omega) $$
for $P$-a.e.\ $\omega \in \Omega$ and this is equal to $\varepsilon_{\min}$ since $-H(c^*\cdot Z_T)$ solves 
$$ \EE \left[L(-X) \mid \F_0\right] \ \longrightarrow \ \essinf $$
by the same arguments as for the maximizer of the expected utility. 
\hfill{$\Box$}

\end{appendix}

%

\end{document}